\documentclass{aa}

\usepackage{graphicx}
\usepackage{txfonts}

\newcommand{\be}{\begin{equation}}
\newcommand{\ee}{\end{equation}}
\newcommand{\ud}{\mathrm{d}}

\begin{document}

\title{Cosmic-ray ionization of molecular clouds}

\author{Marco Padovani\inst{1,2}, Daniele Galli\inst{2} \and Alfred E. Glassgold\inst{3}}

\titlerunning{Cosmic-ray ionization}
\authorrunning{Padovani et al.}

%\offprints{M. Padovani}

\institute{Dipartimento di Astronomia e Scienza dello Spazio,
Universit\`a di Firenze, Largo E. Fermi 2, I-50125 Firenze, Italy\\
\email{padovani@arcetri.astro.it}
\and
INAF-Osservatorio Astrofisico di Arcetri, Largo E. Fermi 5, I-50125 Firenze, Italy\\
\email{galli@arcetri.astro.it}
\and
University of California at Berkeley, Berkeley, CA, 94720 USA
\\
\email{glassgol@berkeley.astro.edu}
}

\abstract
% context heading (optional)
{Low-energy cosmic rays are a fundamental source of ionization for
molecular clouds, influencing their chemical, thermal and dynamical
evolution.}
% aims heading (mandatory)  
{The purpose of this work is to explore the possibility that a
low-energy component of cosmic-rays, not directly measurable from the
Earth, can account for the discrepancy between the ionization rate
measured in diffuse and dense interstellar clouds.}
% methods heading (mandatory)
{We collect the most recent experimental and theoretical data on the
cross sections for the production of H$_2^+$ and He$^+$ by electron and
proton impact, and we discuss the available constraints on the
cosmic-ray fluxes in the local interstellar medium. Starting from
different extrapolations at low energies of the demodulated cosmic-ray
proton and electron spectra, we compute the propagated spectra in
molecular clouds in the continuous slowing-down approximation taking
into account all the relevant energy loss processes.}
% results heading (mandatory)
{The theoretical value of the cosmic-ray ionization rate as a function
of the column density of traversed matter is in agreement with the
observational data only if either the flux of cosmic-ray electrons or
of protons increases at low energies. The most successful models are
characterized by a significant (or even dominant) contribution of the
electron component to the ionization rate, in agreement with previous
suggestions.  However, the large spread of cosmic-ray ionization rates
inferred from chemical models of molecular cloud cores remains to be
explained.}
% conclusions heading
{Available data combined with simple propagation models support the
existence of a low-energy component (below $\sim 100$~MeV) of
cosmic-ray electrons or protons responsible for the ionization of
molecular cloud cores and dense protostellar envelopes.}

\maketitle

\section{Introduction}
\label{introduction}

Cosmic-rays (CRs) play a key role in the chemistry and dynamics of the
interstellar medium (ISM).  First, CR particles are a primary source of
ionization, competing with stellar UV photons (absorbed in a thin layer
of $\sim 4$ magnitudes of visual extinction, McKee~1999) and X-rays
produced by embedded young stellar objects (Krolik \& Kallman~1983;
Silk \& Norman~1983).  The ionization fraction in turn drives the
chemistry of molecular clouds and controls the coupling of the gas with
the Galactic magnetic field (for a good review of the chemistry
that occurs in the ISM in response to CR ionization see
Dalgarno~2006). Second, CRs represent an important source of heating
for molecular clouds because the energy of primary and secondary
electrons produced by the ionization process is in large part converted
into heat by inelastic collisions with ISM atoms and molecules.

In general, the CR ionization rate in the interstellar gas depends on
the relative amount of H, H$_2$ and He (Dalgarno, Yan \& Liu~1999).
The first theoretical determination of the CR ionization rate was
performed for clouds made only by atomic hydrogen by Hayakawa,
Nishimura \& Takayanagi~(1961).  They assumed a proton  specific
intensity (hereafter, for simplicity, {\em spectrum}) proportional to
the proton energy $E_p$ for $0.1~{\rm MeV} < E_p < 10~{\rm MeV}$ and
computed $\zeta^{\rm H}\approx 4\times 10^{-16}$~s$^{-1}$.  Spitzer \&
Tomasko~(1968) determined a value (actually a lower limit) of
$\zeta^{\rm H}\gtrsim 6.8\times 10^{-18}$~s$^{-1}$ for HI clouds,
assuming a CR proton spectrum declining below $E_p\approx 50$~MeV, and
an upper limit of $\zeta^{{\rm H}} \lesssim 1.2\times
10^{-15}$~s$^{-1}$, taking into account an additional flux of $\sim
2$~MeV protons produced by supernova explosions. To obtain the CR
ionization rate of molecular hydrogen, $\zeta^{{\rm H}_2}$, a useful
approximation is $1.5\zeta^{{\rm H}_2}\approx 2.3\zeta^{{\rm H}}$
(Glassgold \& Langer~1974), giving $\zeta^{{\rm H}_2}\approx
10^{-17}$~s$^{-1}$, in agreement with the lower limit on 
$\zeta^{\rm H}$ of Spitzer \& Tomasko~(1968). This value of 
$\zeta^{{\rm H}_2}$ is often referenced as the ``standard'' CR 
ionization rate in molecular clouds.

A major problem in the determination of the CR ionization rate is that
low-energy CRs are prevented from entering the heliosphere by the solar
wind and the interplanetary magnetic field (solar modulation). In
practice, Earth-based measurements of CR fluxes give no information on
the interstellar spectrum of protons and heavy nuclei for energies
below $\sim 1$~GeV/nucleon.  Solar modulation also suppresses the flux
of low-energy CR electrons, that shows considerable fluctuations
already at energies of 10--100~GeV (see e.g. Casadei \& Bindi~2004).
Since the cross section for ionization of molecular hydrogen by
collisions with protons and electrons has a maximum at $\sim 10$~keV
and $\sim 50$~eV, respectively (see Sect.~\ref{reactions_h2}), it is
clear that a knowledge of CR spectrum at low energies is an important
limiting factor for an accurate calculation of the ionization rate in
the ISM. A direct measurement of the shape of the CR spectrum at these
energies will be possible only when spacecrafts such as {\em
Pioneer} and {\em Voyager} are well beyond the heliopause, the
outermost boundary for solar modulation effects, believed to lie at
100--150~AU from the Sun (at present, both Voyagers have already
crossed the solar wind termination shock at 85--95~AU from the Sun).

Over the last three decades, several values of $\zeta^{\rm H}$ ranging
from a few $10^{-17}$~s$^{-1}$ to a few $10^{-16}$~s$^{-1}$ have been
obtained in diffuse interstellar clouds from measurements of the
abundances of various chemical species, in particular OH (Black \&
Dalgarno~1977; Hartquist, Black, \& Dalgarno~1978; Black, Hartquist, \&
Dalgarno~1978) and HD (van Dishoeck \& Black~1986; Federman, Weber \&
Lambert~1996).  However, the derived rates depend sensitively on
several model assumptions, e.g. the value of specific chemical reaction
rates and the intensity of the UV background.  In dense molecular
clouds, the determination of the CR ionization rate is made even more
uncertain by the sensitivity of molecular abundances to the level of
depletion of the various species and the role of small and large grains
in the chemical network. The values of $\zeta^{{\rm H}_2}$ derived by
Caselli et al.~(1998) in a sample of 23 molecular cloud cores (column
density $N({\rm H}_2)\sim 10^{22}$~cm$^{-2}$ ) through DCO$^+$ and
HCO$^+$ abundance ratios span a range of about two orders of magnitudes
from $\sim 10^{-17}$~s$^{-1}$ to $\sim 10^{-15}$~s$^{-1}$, with a
scatter that may in part reflect intrinsic variations of the CR flux
from core to core. Finally, values of $\zeta^{{\rm H}_2}$ of a few
times $10^{-17}$~s$^{-1}$ have been obtained in clouds of higher column
density ($N({\rm H}_2)\sim 10^{23}$--$10^{24}$~cm$^{-2}$) like the
envelopes surrounding massive protostellar sources (van der Tak \& van
Dishoeck~2000; Doty et al.~2002).

The discovery of significant abundances of H$_3^+$ in diffuse clouds
(McCall et al.~1998), confirmed by follow-up detections (Geballe et
al.~1999; McCall et al.~2003; Indriolo et al.~2007), has led to values
of $\zeta^{{\rm H}_2}$ larger by about one order of magnitude than both
the ``standard'' rate and previous estimates based on the abundance of
OH and HD in dense clouds. Given the relative simplicity of the
chemistry of H$_3^+$, it is now believed that diffuse clouds are
characterized by CR ionization rates $\zeta^{{\rm H}_2}\approx 2\times
10^{-16}$~s$^{-1}$ or larger.  This high value of $\zeta^{{\rm H}_2}$
in the diffuse interstellar gas can be reconciled with the lower values
measured in cloud cores and massive protostellar envelopes by
invoking various mechanisms of CR screening in molecular clouds due to
either self-generated Alfv\'en waves in the plasma (Skilling \&
Strong~1976; Hartquist, Doyle \& Dalgarno; Padoan \& Scalo~2005) or to
magnetic mirror effects (Cesarsky \& V\"olk~1978; Chandran~2000).  An
alternative explanation, based on the possible existence of a
low-energy flux of CR particles, is that they can penetrate (and
ionize) diffuse clouds but not dense clouds, as recently proposed by
McCall et al.~(2003; see also Takayanagi~1973 and Umebayashi \&
Nakano~1981). This latter scenario is explored quantitatively in the
present paper.

In this paper, we concentrate on molecular clouds, where hydrogen is
present mostly in molecular form and we can ignore ionization of atomic
hydrogen. In Sect.~\ref{comparison} we then apply our results to
diffuse clouds, where the fraction of hydrogen in molecular form
$f=2N({\rm H}_2)/[N({\rm H}) + 2N({\rm H}_2)]$ has a mean value
$\langle f\rangle \approx 0.6$ (Indriolo et al.~2007), implying that
the column densities of H and H$_2$ are almost equal. This is justified
because the quantity directly measured (or estimated) in the diffuse
clouds examined in Sect.~\ref{comparison} is the ionization rate of
H$_2$ as derived from the measured abundance of H$_3^+$.

The organization of the paper is the following. In
Sect.~\ref{reactions_h2}, \ref{reactions_e} and \ref{reactions_he} we
examine the ionization reactions of CR protons and electrons incident
on H$_2$ and He and other channels of electron production; in
Sect.~\ref{spectra} we discuss the assumed interstellar spectra of CR
protons and electrons; in Sect.~\ref{energy_loss} we discuss the energy
loss mechanisms for CRs; in Sect.~\ref{ionization} we compute the
ionization rate as function of the column density in a cloud; in
Sect.~\ref{comparison} we compare our results with the available
estimates of the CR ionization rate in diffuse and dense clouds;
finally, in Sect.~\ref{conclusions} we summarize our conclusions.

\begin{table}[t]
\begin{center}
\caption{CR reactions in molecular clouds}
\begin{tabular}{lll}
\hline
reaction & cross section & ref. \\
\hline
$p_{\rm CR} + {\rm H}_2 \rightarrow p_{\rm CR} + {\rm H}_2^+ + e$          &  $\sigma_p^{\rm ion.}$        &  \S\ref{p_ion} \\
$p_{\rm CR}+{\rm H}_2 \rightarrow {\rm H} + {\rm H}_2^+$                   &  $\sigma_p^{\rm e.~c.}$       &  \S\ref{p_e_c} \\
$p_{\rm CR} + {\rm H}_2 \rightarrow p_{\rm CR} + {\rm H} + {\rm H}^+ + e$  &  $\sigma_p^{\rm diss.~ion.}$  &  \S\ref{e_diss_ion} \\
$p_{\rm CR} + {\rm H}_2 \rightarrow p_{\rm CR} + 2{\rm H}^+ + 2e$          &  $\sigma_p^{\rm doub.~ion.}$  &  \S\ref{e_double_ion} \\
\hline
$e_{\rm CR} + {\rm H}_2 \rightarrow e_{\rm CR} + {\rm H}_2^+ + e$          &  $\sigma_e^{\rm ion.}$        &  \S\ref{e_ion} \\
$e_{\rm CR} + {\rm H}_2 \rightarrow e_{\rm CR} + {\rm H} + {\rm H}^+ + e$  &  $\sigma_e^{\rm diss.~ion.}$  &  \S\ref{e_diss_ion} \\
$e_{\rm CR} + {\rm H}_2 \rightarrow e_{\rm CR} + 2{\rm H}^+ + 2e$          &  $\sigma_e^{\rm doub.~ion.}$  &  \S\ref{e_double_ion} \\
\hline
$p_{\rm CR} + {\rm He} \rightarrow p_{\rm CR} + {\rm He}^+ + e$            &  $\sigma_p^{\rm ion.}$        &  \S\ref{p_ion_he} \\
$p_{\rm CR}+{\rm He} \rightarrow {\rm H} + {\rm He}+$                      &  $\sigma_p^{\rm e.~c.}$       &  \S\ref{p_e_c_he} \\
\hline
$e_{\rm CR} + {\rm He} \rightarrow e_{\rm CR} + {\rm He}^+ + e$            &  $\sigma_e^{\rm ion.}$        &  \S\ref{e_ion_he} \\
\hline
\end{tabular}
\label{react}
\end{center}
\end{table}

\section{CR reactions with H$_2$}
\label{reactions_h2}

CR particles (electrons, protons, and heavy nuclei) impact with atoms and
molecules of the ISM producing ions and electrons. Table~\ref{react} lists
the main CR ionization reactions involving H$_2$ and He.  In molecular
clouds, a large majority of CR--H$_2$ impacts leads to the formation
of H$_2^+$ via the {\em ionization} reaction
\be
k_{\rm CR} + {\rm H}_2 \rightarrow k_{\rm CR} + {\rm H}_2^+ + e,
\label{ion}
\ee
where $k_{\rm CR}$ is a cosmic-ray particle of species $k$ and energy
$E_k$, with cross section $\sigma_k^{\rm ion.}$.  Here we consider CR
electrons ($k=e$), protons ($k=p$), and heavy nuclei of charge $Ze$
($k=Z$, with $Z \ge 2$).  Low-energy CR protons, in addition, may react
with ambient H$_2$ by {\em electron capture} reactions,
\be
p_{\rm CR}+{\rm H}_2 \rightarrow {\rm H} + {\rm H}_2^+,
\label{p_ec}
\ee
with cross section $\sigma_p^{\rm e.~c.}$.
For an isotropic distribution of CR particles, the production rate of
H$_2^+$ (per H$_2$ molecule) is then
\begin{eqnarray}
\lefteqn{\zeta^{{\rm H}_2} = 4\pi \sum_k
\int_{I({\rm H}_2)}^{E_{\rm max}} j_k(E_k)[1+\phi_k(E_k)]
\sigma_k^{\rm ion.}(E_k)\, E_k} \nonumber \\
& & 
+4\pi\int_{0}^{E_{\rm max}} j_p(E)
\sigma_p^{\rm e.~c.}(E_p)\, E_p, 
\label{zetatot_h2}
\end{eqnarray}
where $j_k(E_k)$ is the number of CR particles of species $k$ per unit
area, time, solid angle and per energy interval (hereafter, we will
refer to $j_k(E_k)$ simply as the spectrum of particle $k$), $I({\rm
H}_2)=15.603$~eV is the ionization potential of H$_2$, and $E_{\rm
max}=10$~GeV is the maximum energy considered. The quantity
$\phi_k(E_k)$ is a correction factor accounting for the ionization of
H$_2$ by secondary electrons. In fact, secondary electrons are
sufficiently energetic to induce further ionizations of H$_2$
molecules, and their relatively short range justifies a local treatment
of their ionizing effects.  The number of secondary ionization produced
per primary ionization of H$_2$ by a particle $k$ is determined by
\be
\phi_k(E_k)\equiv \frac{1}{\sigma_k^{\rm ion.}(E_k)}
\int_{I({\rm H}_2)}^{E^\prime_{\rm max}}P(E_k,E_e^\prime)
\sigma^{\rm ion.}_e(E_e^\prime)\, E_e^\prime,
\ee
where $P(E_k,E_e^\prime)$ is the probability that a secondary electron
of energy $E_e^\prime$ is ejected in a primary ionization by a particle
of energy $E_k$. The spectrum of secondary electrons declines rapidly
with $E_e^\prime$ from the maximum at $E_e^\prime=0$ (Glassgold \&
Langer~1973b; Cecchi-Pestellini \& Aiello~1992).  The function
$\phi_e(E_e)$ giving the number of secondary ionizations after a single
ionization by an electron of energy $E_e$ has been computed by
Glassgold \& Langer~(1973b) for energies of the incident electron up to
10~keV. Above a few 100~eV, $\phi_e$ increases logarithmically with
$E_e$. For secondary electrons produced by impact of particles $k$, we
adopt the scaling $\phi_k(E_k)\approx \phi_e(E_e=m_e E_k/m_k)$ valid in
the Bethe-Born approximation. Calculations by Cravens \&
Dalgarno~(1978) confirm this scaling for protons in the range
1--100~MeV.

In the following subsections we summarize the available data for
the ionization cross sections for proton and electron impact and for
the electron capture cross section.  The ionization of H$_2$ by CR
heavy-nuclei ($Z\ge 2$) is computed in the Bethe-Born approximation as
described in Appendix~\ref{approx_heavy}.

\subsection{Ionization of H$_2$ by proton impact: \\
$p_{\rm CR} + {\rm H}_2 \rightarrow p_{\rm CR} + {\rm H}_2^+ + e$}
\label{p_ion}

\begin{figure}[t]
\begin{center}
\resizebox{\hsize}{!}{\includegraphics{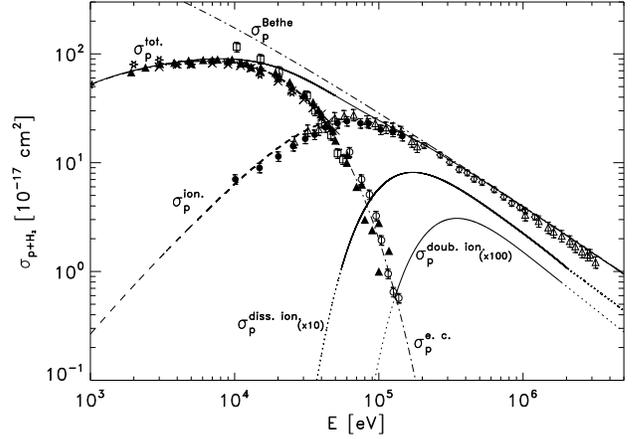}}
\caption[]{Cross sections for proton impact on H$_2$: ionization cross
section $\sigma_p^{\rm ion.}$ (Rudd et al.~1985) and electron capture
$\sigma_p^{\rm e.~c.}$ (Rudd et al.~1983) and total cross section
$\sigma_p^{\rm tot.}$ for production of H$_2^+$. For comparison,
the {\em dot-dashed line} shows the Bethe ionization cross section
multiplied by a factor of 2. The two lower curves show the cross
sections for dissociative ionization and double ionization of H$_2$,
multiplied by a factor of 10 and 100, respectively, obtained from the
corresponding expressions for electron impact at equal velocity.
Experimental data for the ionization cross section:
{\em stars}, Gilbody \& Hasted~(1957);
{\em triangles}, Afrosimov et al.~(1958); 
{\em diamonds}, Hooper et al.~(1961); 
{\em filled circles}, deHeer, Schutten \& Moustafa~(1966);
Experimental data for the electron capture cross section: 
{\em crosses}, Curran, Donahue \& Kasner~(1959);
{\em empty circles}, deHeer, Schutten \& Moustafa~(1966);
{\em asterisks}, McClure~(1966);
{\em squares}, Toburen \& Wilson~(1972).}
\label{Figure1}
\end{center}
\end{figure}

The avalaible experimental data for proton-impact ionization of H$_2$ have
been summarized by Rudd et al.~(1985). The cross section has a maximum
at $E_p\approx 70$~keV and is considerably uncertain below $\sim 1$~keV.
The data were fitted by Rudd et al.~(1985) with expressions appropriate
to the high- and low-energy regions,
\be
\sigma_p^{\rm ion.}=(\sigma_l^{-1}+\sigma_h^{-1})^{-1},
\label{ion_rudd_a}
\ee
where 
\be
\sigma_l=4\pi a_0^2 C x^D, \qquad \sigma_h=4\pi a_0^2[A\ln(1+x)+B]x^{-1},
\label{ion_rudd_b}
\ee
with $x=m_eE_p/m_pI({\rm H})$, $I(\rm H)=13.598$~eV, $A=0.71$,
$B=1.63$, $C=0.51$, $D=1.24$. This expression is compared with
experimental data in Fig.~\ref{Figure1}. For comparison, we also
show in Fig.~\ref{Figure1} the Bethe (1933) cross section for primary
ionization of atomic hydrogen multiplied by a factor of 2. As it is
evident, the Bethe formula reproduces very well the behavior of the
ionization cross section at energies above a few tens of MeV.

\subsection{Ionization of H$_2$ by electron impact: \\
$e_{\rm CR} + {\rm H}_2 \rightarrow e_{\rm CR} + {\rm H}_2^+ + e$}
\label{e_ion}

\begin{figure}[t]
\begin{center}
\resizebox{\hsize}{!}{\includegraphics{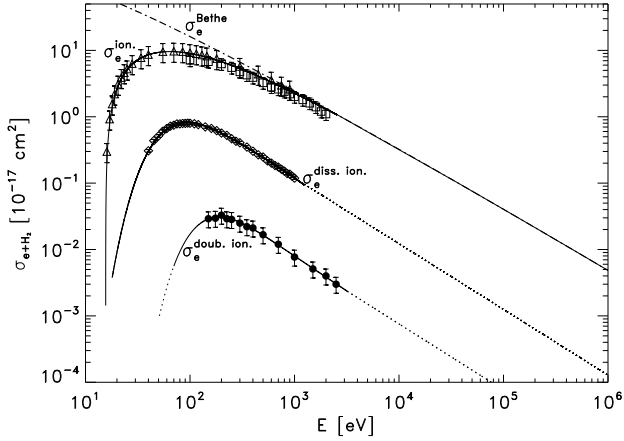}}
\caption[]{Cross sections for electron impact on H$_2$: ionization cross
section $\sigma_p^{\rm ion.}$ (Rudd~1991), dissociative ionization
$\sigma_p^{\rm diss. ion.}$, and double ionization cross section
$\sigma_p^{\rm doub. ion.}$ (polynomial fits of Table~\ref{pol_fit},
{\em solid} part of the curves).
For comparison, the {\em dot-dashed line}
shows the Bethe ionization cross section multiplied by a factor of 2. 
Experimental data for the ionization cross section: 
{\em triangles}, Rapp \& Englander-Golden~(1965); 
{\em squares}, Kossmann, Schwarzkopf \& Schmidt~(1990). 
Experimental data for the dissociative ionization cross section:
{\em diamonds}, Straub et al.~(1996).  
Experimental data for the double ionization cross section:
{\em filled circles}, Kossmann, Schwarzkopf \& Schmidt~(1990).}
\label{Figure2}
\end{center}
\end{figure}

The experimental data for electron-impact ionization of H$_2$ have been
reviewed by Liu \& Shemansky~(2004).  The absolute cross sections for
electron-impact ionization of H$_2$ measured by Straub et al.~(1996) in
the energy range $E_e=17$~eV to $E_e=1$~keV represent the currently
recommended experimental values (Lindsay \& Mangan~2003).  Analytic
expressions and fitting formulae for the ionization cross section have
been derived by Rudd~(1991), Kim \& Rudd~(1994) and Liu \&
Shemansky~(2004).  Here we adopt the semi-empirical model by
Rudd~(1991) that gives an analytical expression valid up to
relativistic velocities based on the theoretical results of
Mott~(1930),
\be
\sigma_e^{\rm ion.}=4\pi a_0^2 N\left[\frac{I({\rm H})}{I({\rm H}_2)}\right]^2
F(t)G(t),
\label{sigma_rudd}
\ee
where $t=E_e/I({\rm H}_2)$, $N=2$ (number of electrons of H$_2$), 
\be
F(t)=\frac{1-t^{1-n}}{n-1}-\left(\frac{2}{1+t}\right)^{n/2}\frac{1-t^{1-n/2}}{n-2},
\label{ft}
\ee
\be
G(t)=\frac{1}{t}\left(A_1\ln t+A_2+\frac{A_3}{t}\right),
\label{gt}
\ee
with $n=2.4\pm 0.2$, $A_1=0.74\pm 0.02$, $A_2=0.87\pm 0.05$,
$A_3=-0.60\pm 0.05$. For comparison, we also show in
Fig.~\ref{Figure2} the Bethe (1933) cross section for primary
ionization of atomic hydrogen multiplied by a factor of 2. The Bethe
formula reproduces very well the behavior of the ionization cross
section at energies above a few tens of keV.

\subsection{Electron capture ionization of H$_2$: \\
$p_{\rm CR} + {\rm H}_2 \rightarrow {\rm H} + {\rm H}_2^+ $}
\label{p_e_c}

In this charge-exchange process, a high-energy CR proton picks up an
electron from the H$_2$ molecule and emerges as a neutral H atom.
The electron capture cross section has been fit by Rudd et al.~(1983)
with the expression
\be
\sigma_p^{\rm e.~c.}=4\pi a_0^2 AN \left[\frac{I({\rm H})}{I({\rm H}_2)}\right]^2
\frac{x^2}{C+x^B+Dx^F},
\label{ec_rudd}
\ee
where $x=E_p/I({\rm H})$, $N=2$ (number of electrons of H$_2$), 
$A=1.044$, $B=2.88$, $C=0.016$, $D=0.136$, $F=5.86$. This
expression is compared in Fig.~\ref{Figure1} with available experimental
results.

\section{Additional reactions of CR electrons and protons with H$_2$}
\label{reactions_e}

Additional ionization reactions that produce electrons are
the {\em dissociative ionization} of H$_2$,
\be
k_{\rm CR} + {\rm H}_2 \rightarrow k_{\rm CR} + {\rm H} + {\rm H}^+ + e,
\label{diss_ion}
\ee
with cross section $\sigma_k^{\rm diss.~ion.}$,
and the {\em double ionization} of H$_2$,
\be
k_{\rm CR} + {\rm H}_2 \rightarrow k_{\rm CR} + 2{\rm H}^+ + 2e.
\label{double_ion}
\ee
with cross section $\sigma_k^{\rm doub.~ion.}$. These two processes contribute 
to the total CR production rate of electrons per H$_2$ molecule,
\begin{eqnarray}
\lefteqn{\zeta^e=4\pi \sum_k
\int_{I({\rm H}_2)}^{E_{\rm max}} j_k(E_k)[1+\phi_k(E_k)]
\sigma_k^{\rm ion.}(E_k)\,\ud E_k} \nonumber \\
& & 
+4\pi \sum_k\int_{E^{\rm diss.~ion.}}^{E_{\rm max}} j_k(E)[1+\phi_k(E_k)]
\sigma_k^{\rm diss.~ion.}(E_k)\,\ud E_k \nonumber \\
& & 
+8\pi \sum_k\int_{E^{\rm doub.~ion.}}^{E_{\rm max}} j_k(E_k)[1+\phi_k(E_k)]
\sigma_k^{\rm doub.~ion.}(E_k)\,\ud E_k.
\label{zetatot_e}
\end{eqnarray}

In the following subsection we examine the cross sections of these two
processes for electron impact reactions, whereas for proton impact we
assume cross sections equal to the corresponding cross sections for
electrons of equal velocity,
\be
\sigma_p^{\rm diss.~ion.}(E_p)=\sigma_e^{\rm diss.~ion.}
(E_e=m_eE_p/m_p)
\ee
and
\be
\sigma_p^{\rm doub.~ion.}(E_p)=\sigma_e^{\rm doub.~ion.}
(E_e=m_eE_p/m_p).
\ee
As shown below, the cross sections of these processes are smaller by at
least one order of magnitude than the corresponding ionization cross
section, and the relative contribution of dissociative ionization and
double ionization to the total electron production rate is expected to
be small.

\subsection{Dissociative ionization of H$_2$ by electron impact:  \\
$e_{\rm CR} + {\rm H}_2 \rightarrow e_{\rm CR} + {\rm H} + {\rm H}^+ + e$}
\label{e_diss_ion}

Absolute partial cross sections for dissociative ionization of H$_2$ by
electron impact (threshold $E^{\rm diss.~ion.}=18.1$~eV) have been
measured by Straub et al.~(1996) for incident electron energies ranging
from $E_e=25$~eV to $E_e=1$~keV (see also Lindsay \& Mangan~2003).
Their results are in agreement with the reanalysis of Van~Zyl \&
Stephen~(1994) of the experimental results of Rapp, Englander-Golden \&
Briglia~(1965), Krishnakumar \& Srivastava~(1994).  For $1~\mbox{keV} <
E_e < 6~\mbox{keV}$, the cross section has been measured by Takayanagi
\& Suzuki (1978). These measurements represent the currently
recommended experimental values (Liu \& Shemansky~2004).  The data of
Straub et al.~(1996) and a polynomial fit of the data are shown in
Fig.~\ref{Figure2}. The coefficients of the polynomial fit, valid for
$18.1~\mbox{eV}<E_e<2$~keV, are given in Table~\ref{pol_fit}.

\begin{table}[t]
\begin{center}
\caption{Fit coefficients for the dissociative ionization and double
ionization cross sections of H$_2$ by electron impact.  The cross
sections are given by $\log(\sigma_e/10^{-18}~\rm{cm}^2)=\sum_n a_n
\log(E_e/\rm{eV})^n$.}
\begin{tabular}{lrr}
\hline
$n$  & $a_n$ (diss.~ion) & $a_n$ (doub.~ion.)  \\
\hline
0    & $-53.23133$ & $-125.8689$ \\
1    & $ 96.57789$ & $ 172.0709$ \\
2    & $-67.57069$ & $-108.8777$ \\
3    & $ 23.32707$ & $ 34.18291$ \\
4    & $-4.004618$ & $-5.358045$ \\
5    & $ 0.272652$ & $ 0.335476$ \\
\hline
\end{tabular}
\label{pol_fit}
\end{center}
\end{table}

\subsection{Double ionization of H$_2$ by electron impact: \\
$e_{\rm CR} + {\rm H}_2 \rightarrow e_{\rm CR} + 2{\rm H}^+ + 2e$}
\label{e_double_ion}

The energy threshold for this reaction is $E^{\rm doub.~ion.}=51$~eV.
The cross section for this reaction is highly uncertain: the
measurements by Edwards et al.~(1988) and Kossmann, Schwarzkopf \&
Schmidt~(1990) disagree by a factor of $\sim 8$. Here we adopt the
latter set of measurements (shown in Fig.~\ref{Figure2}).  The
coefficients of a polynomial fit of these data, valid for
$51~\mbox{eV}<E_e<4$~keV, are given in Table~\ref{pol_fit}.

\section{CR reactions with He}
\label{reactions_he}

The CR production rate of He$^+$ (per He atom) is 
\begin{eqnarray}
\lefteqn{\zeta^{{\rm He}}=4\pi \sum_k
\int_{I({\rm He})}^{E_{\rm max}} j_k(E_k)[1+\phi_k(E_k)]
\sigma_k^{\rm ion.}(E_k)\,\ud E_k} \nonumber \\
& &
+4\pi\int_{0}^{E_{\rm max}} j_p(E)\sigma_p^{\rm e.~c.}(E_p)\,\ud E_p. 
\label{zetatot_he}
\end{eqnarray}
where $I({\rm He})=24.587$~eV is the ionization potential of He,
$\sigma_k^{\rm ion.}$ is the ionization cross sections of He for impact
by particles $k$, and $\sigma_k^{\rm e.~c.}$ is the electron capture
cross section.  In the following subsections we describe the relevant
cross section data proton and electron impact on He.  As in the case of
H$_2$, the ionization of He by CR heavy-nuclei is computed in the
Bethe-Born approximation described in Appendix~\ref{approx_heavy}.

\subsection{Ionization of He by proton impact: \\
$p_{\rm CR} + {\rm He} \rightarrow p_{\rm CR} + {\rm He}^+ + e$}
\label{p_ion_he}

\begin{figure}[t]
\begin{center}
\resizebox{\hsize}{!}{\includegraphics{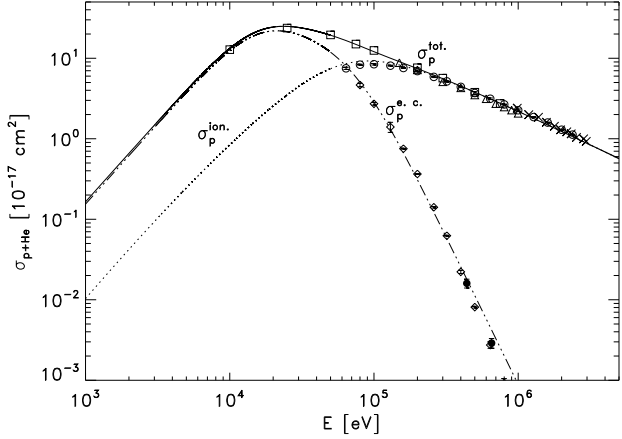}}
\caption[]{Cross sections for proton impact on He: ionization cross
section $\sigma_p^{\rm ion.}$ (Rudd et al.~1983), electron capture
cross section $\sigma_p^{\rm e.~c.}$, and total cross section
$\sigma_p^{\rm tot.}$ for production of He$^+$. Experimental data for
the ionization cross section:
{\rm circles}, Shah \& Gilbody~(1985). 
Data for the electron capture cross section: 
{\em filled circles}, Welsh et al.~(1967); 
{\em diamonds}, Shah \& Gilbody~(1985). 
Data for the total ionization cross section:
{\rm crosses}, Pivovar \& Levchenko~(1967); 
{\em triangles}, Puckett \& Martin~(1970); 
{\rm squares}, DuBois, Toburen \& Rudd~(1984).}
\label{Figure3}
\end{center}
\end{figure}

Experimental measurements of He ionization by proton impact have been
collected and fitted by Rudd et al.~(1985). The cross section has a
maximum at $E_p\approx 100$~keV and is considerably uncertain below
$\sim 10$~keV.  Fig.~\ref{Figure3} shows the available experimental
data.  We adopt the fitting formula of Rudd et al.~(1985) given by
eq.~(\ref{ion_rudd_a}) and (\ref{ion_rudd_b}) with parameters $A=0.49$,
$B=0.62$, $C=0.13$, $D=1.52$.

\subsection{Ionization of He by electron capture: \\
$p_{\rm CR} + {\rm He} \rightarrow {\rm H} + {\rm He}^+$}
\label{p_e_c_he}

The cross section for this charge transfer reaction has been measured
by Welsh et al.~(1967) and Shah \& Gilbody~(1985). The cross section
has a maximum at $E_p\approx 25$~keV, where it is about one order of
magnitude larger than the ionization cross section $\sigma_p^{\rm
ion.}$ (see Fig.~\ref{Figure3}).  Total ionization cross sections
($\sigma_p^{\rm ion.} + \sigma_p^{\rm e.~c.})$ have been reported by
DuBois, Toburen \& Rudd~(1984).

\subsection{Ionization of He by electron impact: \\
$e_{\rm CR} + {\rm He} \rightarrow e_{\rm CR} + {\rm He}^+ + e$}
\label{e_ion_he}

\begin{figure}[t]
\begin{center}
\resizebox{\hsize}{!}{\includegraphics{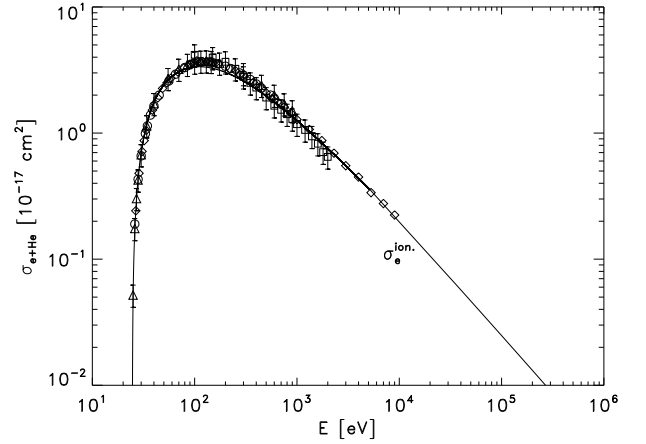}}
\caption[]{Cross section for He ionization by electron impact $\sigma_e^{\rm ion.}$
(Rudd~1991). Experimental data: 
{\em triangles}, Rapp \& Englander-Golden~(1965); 
{\em circles} Montague, Harrison, \& Smith~(1984); 
{\em diamonds}, Shah et al.~(1988);
{\em squares}, Kossmann, Schwarzkopf \& Schmidt~(1990).
}
\label{Figure4}
\end{center}
\end{figure}

Accurate experimental measurements of the cross section of ionization
of He by electron impact are available (see Fig.~\ref{Figure4}) and
are in good agreement with theoretical calculations (Pindzola \&
Robicheaux~2000; Colgan et al.~2006). Here we adopt the fitting formula
of Rudd~(1991) given in eq.~(\ref{sigma_rudd})--(\ref{gt}) with $N=2$,
$n=2.4\pm 0.3$, $A_1=0.85\pm 0.04$, $A_2=0.36\pm 0.09$, and
$A_3=-0.10\pm 0.10$.

\section{Local interstellar spectra}
\label{spectra}

From a theoretical point of view, if one assumes a uniform distribution
(in space and time) of CR sources characterized by a given ``source
spectrum'' (usually a power-law in rigidity), CR propagation models can
generate steady-state local interstellar (LIS) spectra resulting from a
number of processes affecting the CR transport in the Galactic disk,
like nuclear interactions, ionization energy loss, radioactive decay,
escape from the Galaxy, etc.  (see e.g. Berezinsky et al.~1990). These
LIS spectra, in turn, can be used as input for solar modulation
calculations to reproduce the CR spectrum and the relative abundances
of CR particles measured at the Earth.  The LIS spectra obtained in
this way are clearly not uniquely defined, and a considerable range of
LIS spectral shapes can be shown to be consistent with the measured CR
flux with appropriate choices of parameters of the transport model (see
e.g. Mewaldt et al.~2004, especially their Fig.~1).

It is generally assumed that the LIS spectrum characterizes the energy
distribution of CR everywhere in the Galactic disk, as long as the ISM
properties do not depart from the uniform conditions assumed in the
propagation model. With this assumption, Webber~(1998) adopted LIS
spectra for protons and heavy nuclei of energy greater than $10$~MeV
and electrons of energy greater than 2~MeV and combined them with data
from {\em Voyager} and {\em Pioneer} spacecraft measurements out to
60~AU from the Sun to obtain a CR ionization rate $\zeta^{{\rm
H}}\approx 3$--$4\times 10^{-17}$~s$^{-1}$.  This is 5--6 times the
``standard'' rate of Spitzer \& Tomasko~(1968) for atomic hydrogen.

It is very uncertain, however,  whether LIS spectra are really
representative of the whole galactic disk, especially because the Solar
System resides in a low-density ($n\approx 10^{-3}$~cm$^{-3}$) region
produced by $\sim 10$ supernovae exploded over the past $10$~Myr (the
``Local Bubble''). In addition, to compute reliable CR ionization
rates, the demodulated spectra need to be extrapolated down to $\sim
\mbox{keV}$ energies where the ionization cross sections have a maximum
(see Sect.~\ref{reactions_h2}, \ref{reactions_e} and
\ref{reactions_he}).  Given these uncertainties, we discuss in the
remainder of the paper the consequences for the CR ionization rate of
making different assumptions about the low-energy behavior of CR
spectra.  In particular, we consider for both protons and electrons a
``minimum'' and ``maximum'' LIS spectrum compatible with the available
observational constraints, and we compute the resulting ionization
rates with the objective of comparing them with existing data for
diffuse and dense clouds.

\subsection{Proton local interstellar spectrum}

We consider two determinations of the proton LIS spectrum: Webber~(1998,
``minimum'') and Moskalenko et al.~(2002, ``maximum''), labeled
respectively W98 and M02. Their characteristics are the following.

({\em i}\/) W98 estimated the LIS proton spectrum down to $\sim
10$~MeV, starting from an injection spectrum parametrized as a
power-law in rigidity, propagated according to the model of
Webber~(1987) and accounting for solar modulation following
Potgieter~(1995). The effects of solar modulation were refined using
data from the Voyager and Pioneer spacecraft, then at distances of
$\sim 60$--70 AU from the Sun.  The predicted LIS proton spectrum of
W98 has a turnover around $E\approx 100$~MeV because of the dominant
effect of ionization losses at low energies in the Galactic propagation
model. Our extrapolation at low energies is a power-law in energy
with exponent 0.95.

({\em ii}\/) The adopted LIS spectrum of M02 (their ``best-fitting''
case) reproduces the observed spectrum of protons, antiprotons, alphas,
the B/C ratio and the diffuse $\gamma$-ray background. It is obtained
from an injection spectrum which is a double power-law in rigidity with
a steepening below 20 GeV, and a flattening of the diffusion
coefficient below 4 GeV to match the B/C ratio at $E\lesssim 100$~MeV.
At low energies, our extrapolation follows a power-law in energy
with exponent $-1$.

\begin{figure}[t]
\begin{center}
\resizebox{\hsize}{!}{\includegraphics{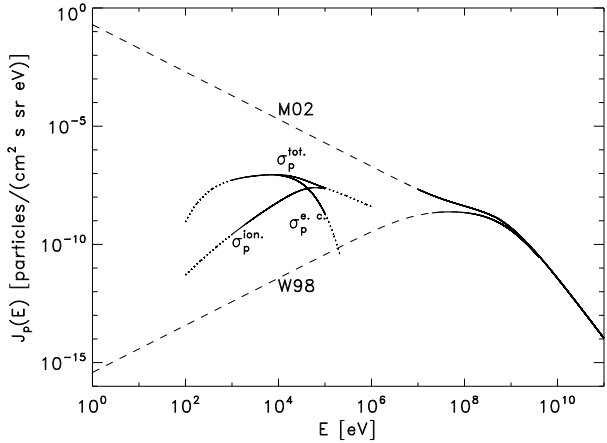}}
\caption[]{Proton LIS spectra of M02 and W98 (upper and lower {\em
solid} curves, respectively). The {\em dashed} curves represent our
power-law extrapolations of the spectra. For comparison, the
cross sections for ionization of H$_2$ by proton impact, electrons
capture, and total ionization are also shown (in arbitrary units).}
\label{Figure5}
\end{center}
\end{figure}

Fig.~\ref{Figure5} shows a comparison of the proton spectrum
according to W98 and M02 (thick lines). The two spectra have been
extrapolated as power laws down to $\sim$~keV energies where the total
ionization cross section, also shown in Fig.~\ref{Figure5}, has a
broad maximum.

\subsection{Electron local interstellar spectrum}

CR electrons (and positrons), although constituting  a small percentage
of the corpuscular radiation, provide important information regarding
interstellar propagation. This happens because CR electrons are more
sensitive probes of ISM conditions than CR nuclei.  In fact, electrons
interact with: ({\em i}\/) the ISM, producing brem\ss trahlung
responsible for the largest part of galactic background at
$\gamma$--frequencies; ({\em ii}\/) radiation fields, generating
radiation by inverse Compton scattering at X- and $\gamma$-frequencies;
({\em iii}\/) magnetic fields, producing synchrotron emission at radio
frequencies.  The electromagnetic radiation emitted by the interaction
of CR electrons with other components of the ISM makes it possible to
establish a relation between the observed radiation spectra and the
energy distribution of the electrons. In particular, observations of
the $\gamma$-ray background in the 10~keV--100~MeV range, combined with
measurements of the Galactic synchrotron spectral index in the
frequency range 10~MHz--10~GHz, provide indirect constraints on the LIS
electron spectrum down to energies of $\sim 100$~MeV. As for the proton
spectrum, we extrapolate the LIS electron spectra to lower energies
with power-laws to reach the peak of the ionization cross section at
$\sim 0.1$~keV. Here we consider two different estimates of the LIS
electron spectrum, both derived by Strong, Moskalenko \&
Reimer~(2000).

({\em i}\/) The first spectrum, labeled C00, corresponds to the
``conventional'' model C of Strong et al.~(2000), and is mostly derived
from radio observations. It reproduces the spectrum of electrons,
protons and alphas above $\sim 10$~GeV, satisfies the limits imposed by
positrons and antiprotons and the constraints on the synchrotron
spectrum, but fails to account for the $\gamma$-ray background,
especially for photon energies below $\sim 30$~MeV and above $\sim
1$~GeV. At low-energies, we have adopted a power-law dependence of
the electron spectrum as $E_e^{0.08}$.

({\em ii}\/) The second spectrum, labeled E00, corresponds to model E00
of Strong et al.~(2000).  It reproduces the $\gamma$ observations at
photon energies below $\sim 30$ MeV by a combination of brem\ss
trahlung and inverse Compton emission, assuming a steepening of the
electron spectrum below $\sim 200$~MeV to compensate for the growth of
ionization losses. The resulting increase in the synchrotron spectrum
occurs at frequencies below 10~MHz, where the radio spectrum decreases
abruptly due to the onset of free-free absorption. To fit OSSE data
would require a LIS electron even steeper than E00, but the excess
$\gamma$ emission at $\sim\mbox{MeV}$ energies may be due to a
population of unresolved point sources (Strong et al.~2000). At
low energies, we have adopted a power-law extrapolation of the spectrum
as $E_e^{-1}$.

\begin{figure}[t]
\begin{center}
\resizebox{\hsize}{!}{\includegraphics{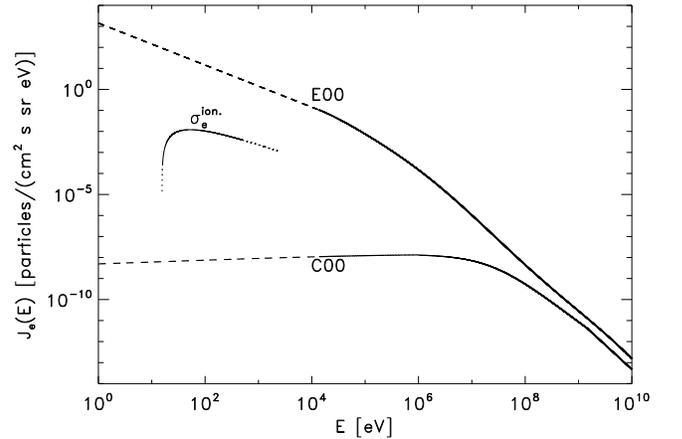}}
\caption[]{Electron LIS spectra of E00 and C00 (upper and lower {\em
solid} curves, respectively). The {\em dashed} curves represent our
extrapolations of the spectra. For comparison, the cross section
for ionization section of H$_2$ by electron impact is also shown (in
arbitrary units).}
\label{Figure6}
\end{center}
\end{figure}

In Fig.~\ref{Figure6} we compare the two LIS electron spectra E00
and C00 assumed in this work.

\subsection{CR ionization rate for the local interstellar spectra}

The values of $\zeta_k^{{\rm H}_2}$, $\zeta_k^e$ and $\zeta_k^{{\rm
He}}$ per H$_2$ molecule and He atom, respectively, obtained from
numerical integration of eq.~(\ref{zetatot_h2}), (\ref{zetatot_e}) and (\ref{zetatot_he}), with the $j_k(E_k)$ taken to be the adopted LIS
spectra, are listed in Tab.~\ref{ion_LIS}. We have assumed a mixture of
H$_2$ and He with $f_{{\rm H}_2}=0.83$ and $f_{\rm He}=0.17$,
corresponding to a He/H ratio of 0.1.  We also list in
Table~\ref{ion_LIS} the energy density of each CR component, defined as
\be 
{\cal E}_k=4\pi\int_0^\infty \frac{j_k(E_k)E_k}{v_k(E_k)}\,\ud E_k
\label{en_den}
\ee
where $j_k(E_k)$ is the particle's LIS spectrum and
$v_k(E_k)=c(E_k^2/m_k^2c^4+2E_k/m_kc^2)^{1/2}/(1+E_k/m_kc^2)$ is the
velocity of particle $k$ with kinetic energy $E_k$. We compute the
total energy density of CR as $\sum_k{\cal E}_k\approx (1+\xi) {\cal
E}_p$, where $\xi=0.41$ is the correction factor for the abundance of
He and heavy nuclei (see Appendix~\ref{approx_heavy}). The results
listed in Table~\ref{ion_LIS} suggest the following considerations:

({\em i}\/) Protons and heavy nuclei (plus secondary electrons) can
produce ionization rates ranging from $\sim 10^{-17}$~s$^{-1}$ 
(in the case of the the spectrum W98, decreasing below $E_p\approx 100$~MeV)
to $\sim 10^{-14}$~s$^{-1}$ (spectrum M02, increasing below
$E_p\approx 100$~MeV).  The contribution of CR electrons to the
ionization rate is negligible if the LIS electron spectrum flattens
below $E_e\approx 10$~MeV (spectrum C00), but can become dominant if
the spectrum increases at low energies.  In practice, the ionization
rate is proportional to the flux of CR particles in the energy range
where the contribution to the integrals in eq.~(\ref{zetatot_h2}),
(\ref{zetatot_e}) and (\ref{zetatot_he}) is larger (see
Sect.~\ref{ionization} and Fig.~\ref{Figure14}).

({\em ii}\/) The ratio of the CR ionization rate of He and H$_2$
depends on the shape and absolute value of the assumed spectra. For CR
protons, the ratio $\zeta_p^{{\rm He}}/\zeta_p^{{\rm H}_2}$ varies
between 0.15 (spectrum M02) and 0.64 (spectrum W98), whereas for
electrons it varies between 0.38 (spectrum E00) and 0.65 (spectrum
C00).  In general, since the ionization cross section for He decreases
faster than that of H$_2$ below the maximum, CR spectra rising with
decreasing energy result in a lower value of $\zeta^{\rm
He}/\zeta^{{\rm H}_2}$. Given the sensitivity of modeled steady-state
abundances of species like C, O$_2$, H$_2$O, H$_3^+$ in dense clouds to
the value of $\zeta^{\rm He}/\zeta^{{\rm H}_2}$ (Wakelam et al.~2006),
it might be possible to constrain this ratio from a careful combination
of molecular line observations and chemical model predictions.

({\em iii}\/) As anticipated, the CR production rate of electrons in
molecular clouds $\zeta_k^e$ is dominated by the CR ionization of H$_2$
(Sect.~\ref{reactions_h2}) and He (Sect.~\ref{reactions_he}).  The
contributions of dissociative ionization and double ionization to
$\zeta_k^e$ are small, about 5.5\% and 0.32\% of the rate of production
of electrons by single ionization of H$_2$, respectively, independent
of the adopted spectrum.

({\em iv}\/) The production rate of electrons, $\zeta_k^e$, is
generally larger than (but close to) the production rate of H$_2^+$.
For the W98 proton spectrum, the C00 and E00 electron spectra,
$\zeta_k^{{\rm H}_2} \approx 0.83$--0.87$\zeta_k^e$. However, since we
have included in the expression for $\zeta^{{\rm H}_2}$ the electron
capture reaction (\ref{p_ec}) whose cross section peaks at a lower
energy than the ionization reaction (\ref{ion}) as shown in
Fig.~\ref{Figure1}, a CR proton spectrum rising at low energies may
result in $\zeta_p^{{\rm H}_2} > \zeta_p^e$, as in the case of the M02
spectrum.

({\em v}\/) With our assumed LIS spectra, the total CR energy density
varies from a minimum of $0.970$~eV~cm$^{-3}$ (W98 plus C00) and a
maximum of $1.80$~eV~cm$^{-3}$ (M02 plus E00), corresponding to an
equipartition magnetic field of $6.2$~$\mu$G and $8.5$~$\mu$G,
respectively.  These equipartition values are compatible with the
``standard'' value of the magnetic field of $6.0\pm 1.8$~$\mu$G in the
cold neutral medium of the Galaxy (Heiles \& Troland~2005). They have
interesting consequences for the location of the solar wind termination
shock (see discussion in Webber~1998).

\begin{table}[t]
\begin{center}
\caption{CR ionization rates $\zeta^{{\rm H}_2}_k$ 
and $\zeta^{\rm He}_k$ (per H$_2$ and per He, respectively),
electron production rate $\zeta^e_k$, and energy densities ${\cal E}_k$ 
of CR protons ($p$) and  electrons ($e$) 
for the LIS spectra assumed in this work. The proton ionization rates include
the contribution of heavy nuclei.}
\begin{tabular}{lllllll}
\hline
$k$   &  ref.      & $\zeta_k^{{\rm H}_2}$   & $\zeta_k^{\rm He}$     & $\zeta_k^e$  & ${\cal E}_k$   \\                &            & (s$^{-1}$)              & (s$^{-1}$)             & (s$^{-1}$)   & (eV~cm$^{-3}$) \\
\hline
$p$ &      W98   & $2.08\times 10^{-17}$   & $1.33\times 10^{-17}$  & $2.50\times 10^{-17}$ & 0.953 \\
$p$ &      M02   & $1.48\times 10^{-14}$   & $2.16\times 10^{-15}$  & $3.49\times 10^{-15}$ & 1.23  \\
\hline
$e$   &    C00   & $1.62\times 10^{-19}$   & $1.05\times 10^{-19}$  & $1.94\times 10^{-19}$ & 0.0167 \\
$e$   &    E00   & $6.53\times 10^{-12}$   & $2.46\times 10^{-12}$  & $7.45\times 10^{-12}$ & 0.571 \\
\hline
\end{tabular}
\label{ion_LIS}
\end{center}
\end{table}

It is important to stress that the CR ionization rates listed in
Table~\ref{ion_LIS} have been obtained by integrating the spectra and
the cross sections down to the ionization threshold of H$_2$ and He,
and they must therefore be considered as upper limits on the ionization
rate. This is especially true for the electron spectrum E00, which
results in ionization rates exceeding the observed values by more than
three orders of magnitude (see Sect.~\ref{comparison}).  In the past,
LIS spectra have been used to compute the CR ionization rate in the ISM
assuming an appropriate lower cut-off in the CR energy (e.g. Nath \&
Biermann~1994; Webber~1998).  In this work, we use the LIS spectra to
define the energy distribution of CR particles incident on the surface
of the cloud. As we show in Sect.~\ref{energy_loss} and
\ref{ionization}, the low-energy tail of the CR spectrum is strongly
(and rapidly) modified by various energy loss processes when the
particles propagate in a medium denser than the local ISM.

\section{Energy losses of CRs in the ISM}
\label{energy_loss}

The penetration of primary CR and secondary particles in interstellar
clouds was studied by Takayanagi~(1973) and more in detail by
Umebayashi \& Nakano~(1981). In this paper we adopt the LIS spectra
discussed in Sect.~\ref{spectra} to characterize the incident spectra
and we follow the propagation of CR particles inside a molecular cloud
with the so-called {\em continuous-slowing-down approximation}
(hereafter CSDA) \footnote{In the astrophysical literature this
approximation is often referred to as the {\em continuous energy loss
regime}, and when the propagation is dominated by these losses, it is
often known as the {\em thick target approximation} (see e.g. Ramaty \&
Lingenfelter~1975; Ramaty, Kozlovsky \& Lingenfelter~1996).}.  In this
approximation, the ``degradation spectrum'' of the CR component $k$
resulting from the energy loss of the incident particles and the
generation of secondary particles is proportional to the inverse of the
{\em energy loss function}, defined by
\be
L_k(E_k)=-\frac{1}{n({\rm H}_2)}\left(\frac{\ud E_k}{\ud \ell}\right),
\label{eloss}
\ee
where $n({\rm H}_2)$ is the density of the medium in which the
particles propagate and $\ell$ is the path length. Since we consider
only energy losses in collisions with H$_2$, our results are applicable
only to clouds in which hydrogen is mostly in molecular form.

In the following we consider CR propagation in molecular clouds
assuming a plane-parallel geometry. It is convenient to introduce the
column density of molecular hydrogen $N({\rm H}_2)$,
\be
N({\rm H}_2)=\int n({\rm H}_2)\,\ud \ell,
\ee
and to rewrite the energy loss function (eq.~\ref{eloss}) as
\be
L_k(E_k)=-\frac{\ud E_k}{\ud N({\rm H}_2)}.
\label{L(E)N}
\ee
Let us then define $j_k(E_k,N)$ as the spectrum of CR particles of
species $k$ at depth $N({\rm H}_2)$, with $j_k(E_k,0)$ representing
the LIS spectrum incident on the cloud's surface, defined by a column
density $N({\rm H}_2)=0$. To compute $j_k(E_k,N)$ we must consider all
the processes that degrade the energy of the incident CR particles.
Assuming that the direction of propagation does not change significantly
inside the cloud, it follows from eq.~(\ref{L(E)N}) that particles of
initial energy $E_{k,0}$ reach energy $E_k<E_{k,0}$ as a consequence of
energy losses after propagating across a column density $N({\rm H}_2)$
given by
\be
N({\rm H}_2)=-\int_{E_{k,0}}^{E_k}\frac{\ud E_k}{L_k(E_k)} 
=n({\rm H}_2)[R_k(E_{k,0})-R_k(E_k)],
\label{defN}
\ee
where $R_k(E_k)$ is the {\em range}, defined as
\be
R_k(E_k)=\int_{E_k}^0\ud \ell=
\int_0^{E_k}\frac{\ud E_k}{-(\ud E_k/\ud x)}=
\frac{1}{n({\rm H}_2)}\int_0^{E_k}\frac{\ud E_k}{L_k(E_k)}.
\label{defrange}
\ee

Conservation of the the number of CR particles of each species implies 
\be
j_k(E_k,N)\,\ud E_k=j(E_{k,0},0)\,\ud E_{k,0},
\ee
where, for a given value of $N({\rm H}_2)$, the infinitesimal variation $\ud
E_{k,0}$ of the particle's initial energy corresponds to an infinitesimal
variation $\ud E_k$ of its energy at a depth $N({\rm H}_2)$ given by
\be
\frac{\ud E_k}{L_k(E_k)}=\frac{\ud E_{k,0}}{L_k(E_{k,0})}
\label{jacob}
\ee
(we ignore here that electron capture reactions of CR protons with H$_2$ and He
do not conserve the number of CR protons).
Thus, the relation between the incident spectrum $j_k(E_{k,0})$ and the
spectrum at depth $j_k(E_k,N)$ in the CSDA is
\be
j_k(E_k,N)=j_k(E_{k,0})\frac{\ud E_k}{\ud E_{k,0}}=
j_k(E_{k,0})\frac{L_k(E_{k,0})}{L_k(E_k)}.
\label{jE0jE}
\ee

The energy loss functions for electrons and protons in H$_2$ are shown
in Fig.~\ref{Figure7}.  Some energy loss processes are common to CR
protons and electrons, like Coulomb interactions, inelastic collisions
and ionization; others are peculiar to protons (elastic collisions,
pion production and spallation), others to electrons (brem\ss trahlung,
synchrotron emission and inverse Compton scattering). These processes
are briefly reviewed in the following subsections.

\subsection{Energy loss of protons colliding with H$_2$}
\label{energy_loss_p}

To determine the energy loss function of protons we have used the
results collected by Phelps~(1990) for energies in the range from
10$^{-1}$~eV to 10$^4$~eV. For higher energies, between 1~keV and
10~GeV, we have used data from the NIST Database\footnote{\tt
http://physics.nist.gov/PhysRefData/Star/Text} for atomic hydrogen
multiplied by a factor of 2 to obtain the corresponding values for
collisions with molecular hydrogen (NIST data do not include pion
production at energies higher than about $0.5$~GeV, that we computed
following Schlickeiser~2002). The resulting energy loss function is
shown in Fig.~\ref{Figure7}. The broad peak in $L_p(E_p)$ at
$E_p\approx 10$~eV is due to elastic collisions and to the
excitation of rotational and vibrational levels, the peak at
$E_p\approx 100$~keV to ionization, and the rapid increase at energies
above $\sim 1$~GeV is due to pion production.  For the low ionization
levels characteristic of molecular clouds, the energy loss for Coulomb
interactions of CRs with ambient electrons can be neglected at energies
above $\sim 1$~eV (dashed line in Fig.~\ref{Figure7}).

In Fig.~\ref{Figure8} we show the quantity $n({\rm H}_2)R_p(E_p)$, obtained
with a numerical integration of eq.~(\ref{defrange}), compared with data
from the NIST Database at energies from 1 keV to 10 GeV. We also show the
fit adopted by Takayanagi~(1973) in a limited range of energies and the
results of Cravens \& Dalgarno~(1978). As one can see, except for energies
higher than $\sim 100$~MeV, where the NIST data do not include energy
losses by pion production, the agreement between our results and the
NIST data is very good.

\subsection{Energy loss of electrons colliding with H$_2$}
\label{energy_loss_e}

\begin{figure}[t]
\begin{center}
\resizebox{\hsize}{!}{\includegraphics{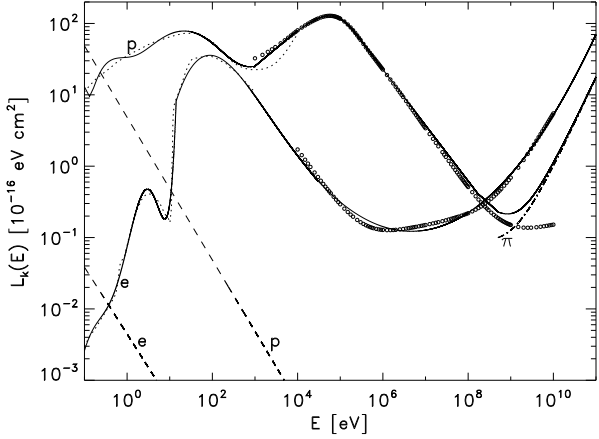}}
\caption[]{Energy loss functions $L_e(E_e)$ and $L_p(E_p)$ for electrons
and protons colliding with H$_2$ ({\em solid} curves), compared with
NIST data ({\em circles}); {\em dashed} curves show Coulomb losses for 
a fractional electron abundance $n(e)/n({\rm H}_2)=10^{-7}$; {\em
dash-dotted} curves labelled with $\pi$ represent the energy loss by pion
production computed following Schlickeiser~(2002); {\em dotted} curves show
the results by Phelps~(1990) and Dalgarno et al.~(1999) for $p$--H$_2$
and $e$--H$_2$, respectively.}
\label{Figure7}
\end{center}
\end{figure}

\begin{figure}[t]
\begin{center}
\resizebox{\hsize}{!}{\includegraphics{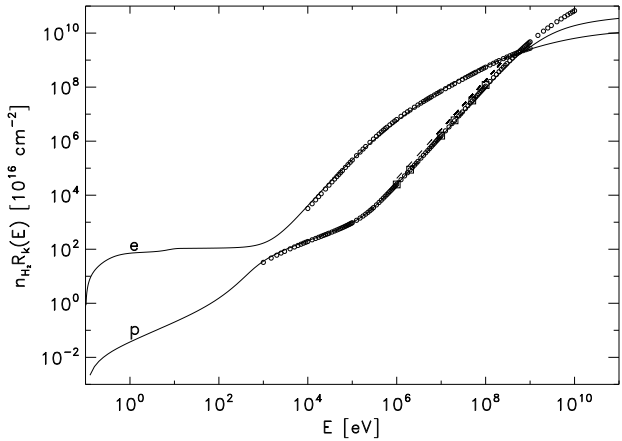}}
\caption[]{Range $R_e(E_e)$ and $R_p(E_p)$ for electrons and protons
colliding with H$_2$ ({\em solid} curves), compared with NIST data ({\em
circles}) and the results of Cravens \& Dalgarno~(1978, {\em squares});
the {\em dashed} curve shows the fit by Takayanagi~(1973)}.
\label{Figure8}
\end{center}
\end{figure}

To determine the electron energy loss function we have adopted the
results of Dalgarno et al.~(1999) for $10^{-2}~{\rm eV} \le E_e \le
1~{\rm keV}$ and those of Cravens, Victor \& Dalgarno~(1975) for
$1~{\rm eV} \le E_e \le 10$~keV. For higher energies, $10~{\rm keV} \le
E_e \le 10~{\rm GeV}$, we have adopted the loss function for electron-H
collisions from the NIST Database multiplied by a factor of 2. The
resulting energy loss function is also shown in
Fig.~\ref{Figure7}.  The first peak in $L_e(E_e)$ is due to the
excitation of vibrational levels, the second to the excitation of the
electronic levels and ionization, while at higher energies the energy
loss function is dominated by brem\ss trahlung.  As in the case of CR
protons, we can neglect the contribution of Coulomb interactions for
electrons at energies above $\sim 1$~eV.  In Fig.~\ref{Figure8}, we show
the range for electrons in H$_2$, obtained as in the case of CR
protons, compared with data from the NIST Database for $10~{\rm keV}
\le E_e \le 1~{\rm GeV}$.

\section{CR ionization rate in diffuse and dense clouds}
\label{ionization}

To compute the CR ionization rate in the ISM as a function of the column
density $N({\rm H}_2)$ of traversed matter, we follow the method of
Takayanagi~(1973). First, varying $E_k$ and $E_{k,0}$ from 0.1 eV to
100 GeV, we determine the column density from the difference between
$R_k(E_{k,0})$ and $R_k(E_k)$.  Second, tracing the level contours of
the surface $N(E_{k,0},E_k)$ at different values of $N({\rm H}_2)$,
we obtain the relation between the energy of the incident CR particle,
$E_{k,0}$, and the residual energy $E_k$, when the particle has covered
a path inside the cloud corresponding to a given value of the column
density. We then fit the resulting $E_{k,0}$ vs. $E_k$ relation at fixed
$N({\rm H}_2)$ with the expression
\be
E_{k,0}(E_k,N)=\left(cE_k^b+\frac{N}{N_0}\right)^{1/b},
\label{abc}
\ee
\noindent
where $E_k$ and $E_{k,0}$ are in eV, $N$ and $N_0$ in cm$^{-2}$,
$b$ and $c$ are non-dimensional.

\begin{figure}[t]
\begin{center}
\resizebox{\hsize}{!}{\includegraphics{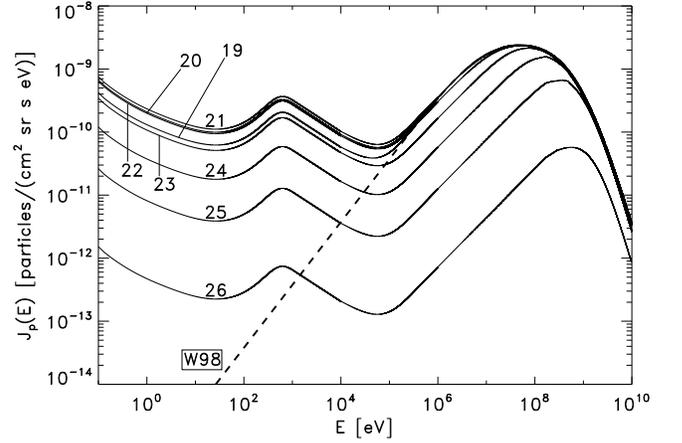}} 
\caption[]{{\em Dashed} curve: LIS proton spectrum W98 incident on the
cloud's surface; {\em dash-dotted} curves: attenuated proton spectra
at increasing depth in the cloud labeled by values of $\log[N({\rm
H}_2)/{\rm cm}^{-2}]$.}
\label{Figure9}
\end{center}
\end{figure}

\begin{figure}[t]
\begin{center}
\resizebox{\hsize}{!}{\includegraphics{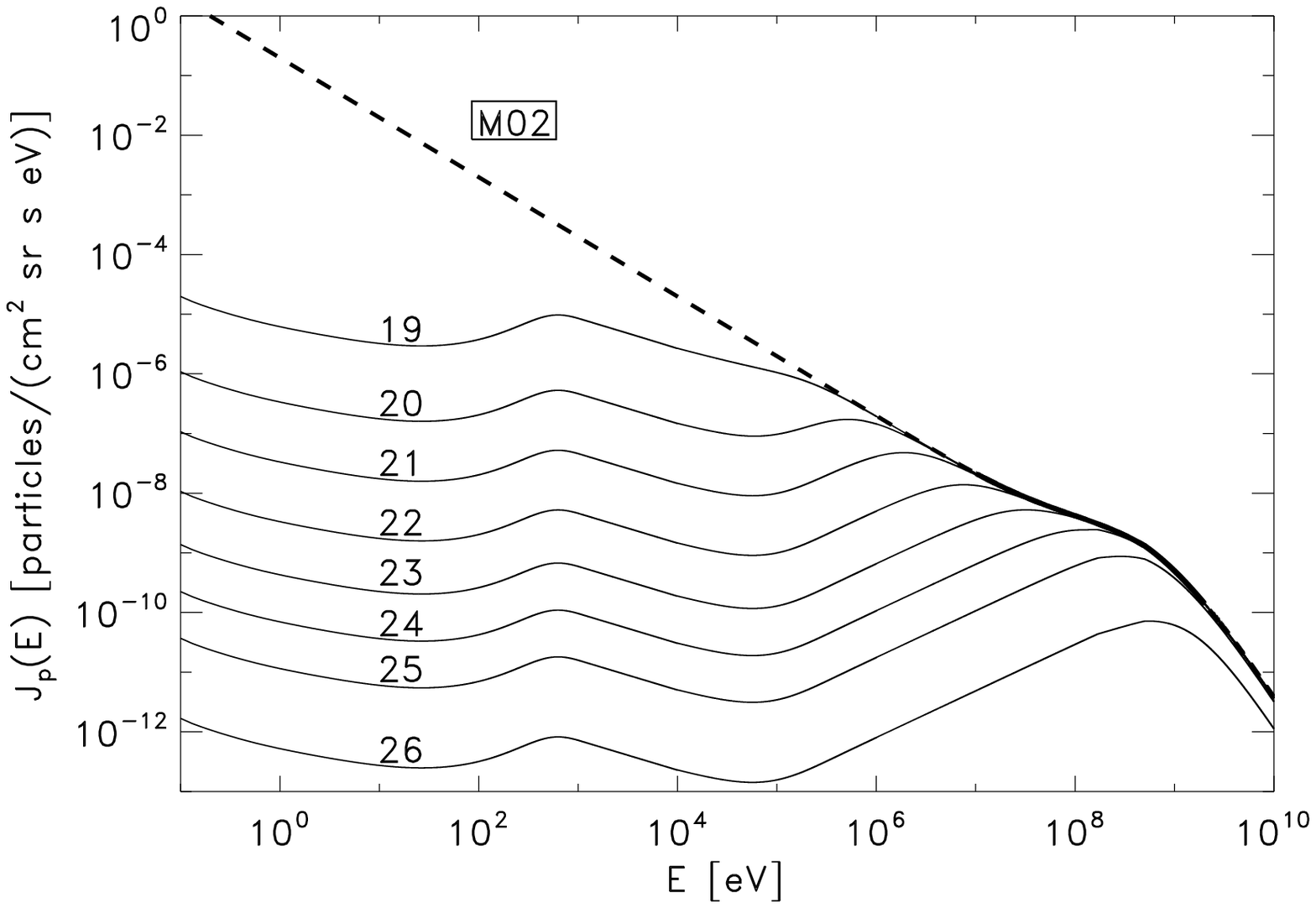}}
\caption[]{{\em Dashed} curve: LIS proton spectrum M02 incident on the
cloud's surface; {\em dash-dotted} curves: attenuated proton spectra
at increasing depth in the cloud labeled by values of $\log[N({\rm
H}_2)/{\rm cm}^{-2}]$.}
\label{Figure10}
\end{center}
\end{figure}

\begin{figure}[t]
\begin{center}
\resizebox{\hsize}{!}{\includegraphics{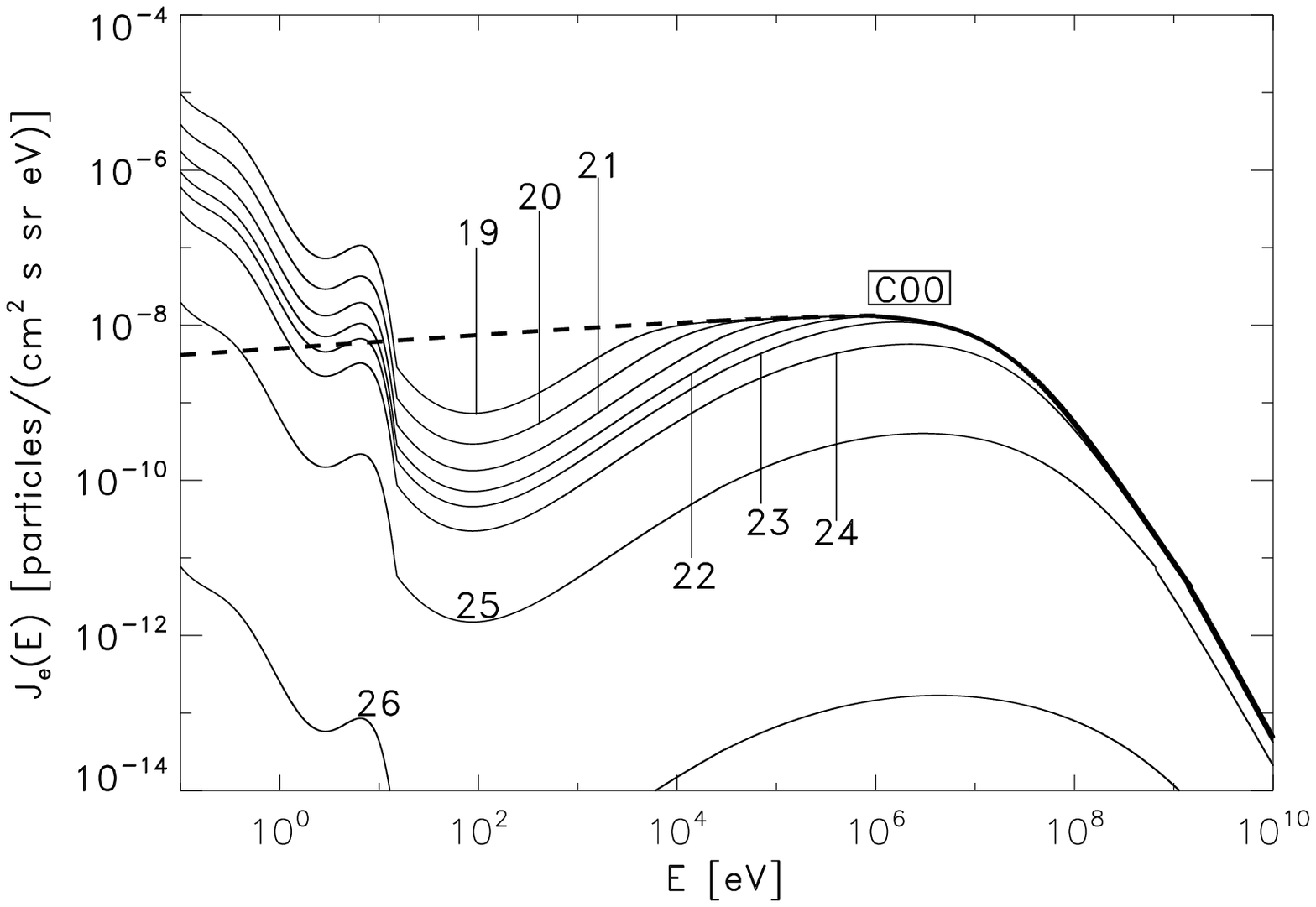}} 
\caption[]{{\em Dashed} curve: LIS electron spectrum C00 incident on the
cloud's surface; {\em dash-dotted} curves: attenuated proton spectra
at increasing depth in the cloud labeled by values of $\log[N({\rm
H}_2)/{\rm cm}^{-2}]$.}
\label{Figure11}
\end{center}
\end{figure}

\begin{figure}[t]
\begin{center}
\resizebox{\hsize}{!}{\includegraphics{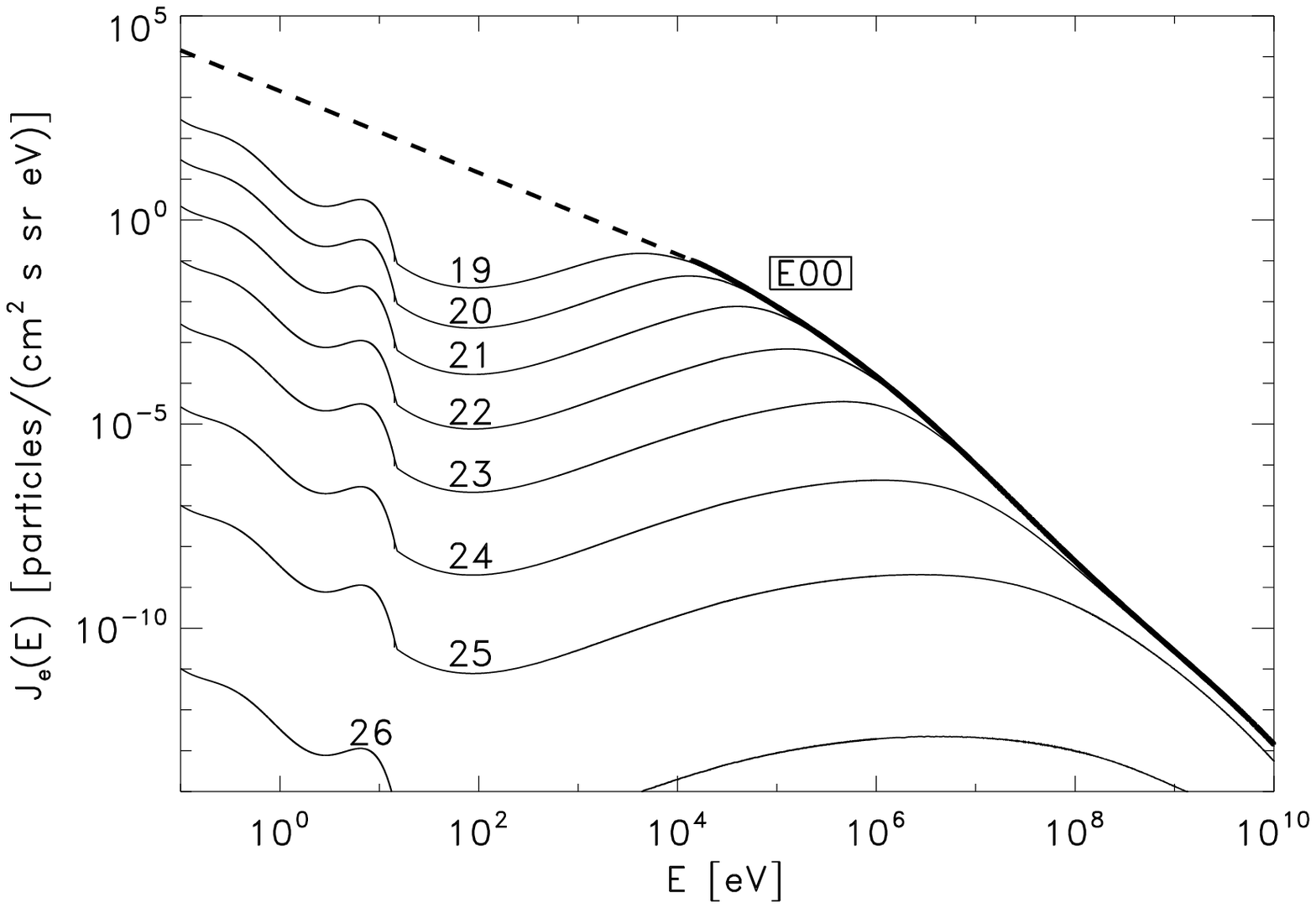}}
\caption[]{{\em Dashed} curve: LIS electron spectrum E00 incident on the
cloud's surface; {\em dash-dotted} curves: attenuated proton spectra
at increasing depth in the cloud labeled by values of $\log[N({\rm
H}_2)/{\rm cm}^{-2}]$.}
\label{Figure12}
\end{center}
\end{figure}

In Fig.~\ref{Figure9}, \ref{Figure10}, \ref{Figure11}, and \ref{Figure12} we
show the CR spectrum obtained from eq.~(\ref{jE0jE}) and (\ref{abc})
for protons and electrons at values of $N({\rm H}_2)$ ranging from
$10^{19}$~cm$^{-2}$ to $10^{26}$~cm$^{-2}$, inside a molecular cloud
for the two incident spectra of protons and electrons described in
Sect.~\ref{spectra}.  One can notice the correspondence between the
shape of the proton spectra shown in Fig.~\ref{Figure9} and
\ref{Figure10}, and the energy loss function $L_p(E)$ shown in
Fig.~\ref{Figure7}. In fact, the relative minimum at about $10$~eV
in the attenuated spectrum corresponds to the energy loss peak due to
elastic interactions and excitation of roto-vibrational levels, and the
minimum at about $100$~keV corresponds to the energy loss peak due to
ionization. The same correspondence can be seen between electron
spectra (Fig.~\ref{Figure11} and \ref{Figure12}) and the energy loss
function $L_e(E_e)$ (Fig.~\ref{Figure7}): the minima in the
spectrum at about 1 eV and 100~eV are caused by the energy loss due to
the excitation of vibrational levels, and to the excitation of
electronic levels and ionization, respectively. This is a well-known
property of the CSDA, where one approximately obtains
$j_k(E_k,N)\propto 1/L_k(E_k)$ independent on the column density if
$N({\rm H}_2)\ll N_0$ (see eq.~\ref{abc}).

\begin{figure}[t]
\begin{center}
\resizebox{\hsize}{!}{\includegraphics{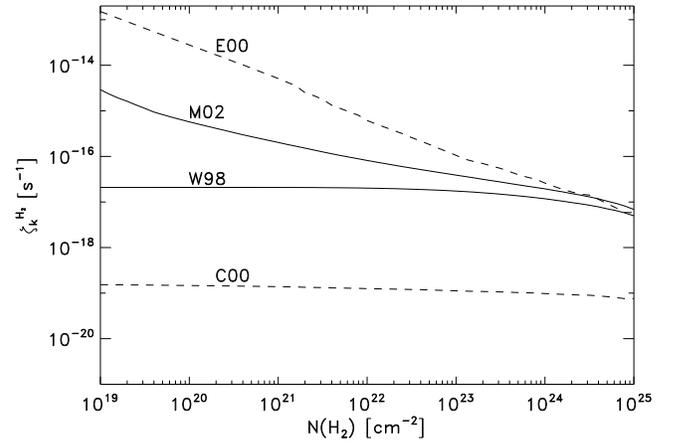}} 
\caption[]{CR ionization rate $\zeta_k^{{\rm H}_2}$ as a function of the
column density $N({\rm H}_2)$. {\em Solid} curves: contribution of 
CR protons (spectra W98 and M02); {\em dashed} curves, contribution 
of CR electrons (spectra C00 and E00).}
\label{Figure13}
\end{center}
\end{figure}

We are now able to calculate the CR ionization rate inside a molecular
cloud as a function of the column density, with the attenuated spectra
given by eq.~(\ref{jE0jE}). We compute the CR ionization rate for
$N({\rm H}_2)$ between $10^{19}$~cm$^{-2}$ and $10^{25}$~cm$^{-2}$, and we
show the results for the four incident LIS spectra in Fig.~\ref{Figure13}.

As a result of the detailed treatment of CR propagation, the
decrease of the ionization rate with increasing penetration in the
cloud at column densities in the range $\sim
10^{20}$--$10^{25}$~cm$^{-2}$ is characterized by a power-law behavior,
rather than exponential attenuation, and can be approximated as
\be
\zeta_k^{{\rm H}_2}\approx 
\zeta_{0,k}\left[\frac{N({\rm H}_2)}{10^{20}~\mbox{cm$^{-2}$}}\right]^{-a}.
\label{zetaNep}
\ee
We have fitted this expression to the numerical results shown in
Fig.~\ref{Figure13}. The coefficients $\zeta_{0,k}$ and $a$ are given in
Table~\ref{tabNep}. The exponential attenuation of the CR ionization
rate sets in for column densities larger than $\sim 10^{25}$~cm$^{-2}$,
where $\zeta_k^{{\rm H}_2}$ depends essentially on the flux of CR
particles in the high-energy tail of the incident spectrum (above $\sim
0.1$--1~GeV), and directly measurable on the Earth. In this regime, the
attenuation of the CR ionization rate is expressed as function of the
surface density of traversed matter $\Sigma=\mu m_p N({\rm H}_2)$,
where $m_p$ is the proton mass and $\mu=2.36$ is the molecular weight
for the assumed fractional abundances of H$_2$ and He ($f_{{\rm
H}_2}=0.82$ and $f_{\rm He}=0.18$). For $\Sigma \gtrsim 1$~g~cm$^{-2}$,
we can fit the CR ionization rate as
\be
\zeta_k^{{\rm H}_2}\approx
\zeta_{0,k}\exp\left(-\frac{\Sigma}{\Sigma_{0,k}}\right)
\label{zetachi}
\ee
where $\Sigma_{0,k}$ is the attenuation surface density. In
Table~\ref{tabchi} we list the values of $\zeta_{0,k}$ and
$\Sigma_{0,k}$ obtained with the four spectra considered in this work.
The values for the attenuation surface density $\Sigma_{0,p}$ listed in
Table~\ref{tabchi} are significantly lower than the ``standard'' value
of Nakano \& Tademaru~(1972) and Umebayashi \& Nakano~(1981), who
obtain $\Sigma_{0,p}\simeq 96$~g~cm$^{-2}$ for $\Sigma\gtrsim
50$~g~cm$^{-2}$ (see also Umebayashi \& Nakano~2009).

\begin{table}[t]
\begin{center}
\caption{Fitting coefficients for eq.~(\ref{zetaNep}), describing the
attenuation of the CR ionization rate for protons ($p$, also including
heavy nuclei) and electrons ($e$). Eq.~\ref{zetaNep}) is valid in the
column density range $10^{20}~\mbox{cm$^{-2}$}\lesssim N({\rm
H}_2) \lesssim 10^{25}~\mbox{cm$^{-2}$}$.}
\begin{tabular}{llll}
\hline
$k$ & spectrum   & $\zeta_{0,k}$        & $a$  \\
    &            & (s$^{-1}$)           &      \\
\hline
$p$ & W98        & $2.0\times 10^{-17}$ & $0.021$ \\
$p$ & M02        & $6.8\times 10^{-16}$ & $0.423$ \\
\hline
$e$   & C00      & $1.4\times 10^{-19}$ & $0.040$ \\
$e$   & E00      & $2.6\times 10^{-14}$ & $0.805$ \\
\hline
\end{tabular}
\label{tabNep}
\end{center}
\end{table}

\begin{table}[t]
\begin{center}
\caption{Fitting coefficients for eq.~(\ref{zetachi}) describing the 
attenuation of CR protons ($p$, also including heavy nuclei) and electrons ($e$).
Eq.~\ref{zetachi}) is valid for $N({\rm H}_2)\gtrsim 10^{25}~\mbox{cm$^{-2}$}$.}
\begin{tabular}{llll}
\hline
$k$ & spectrum   & $\zeta_{0,k}$        & $\Sigma_{0,k}$  \\
    &                & (s$^{-1}$)           & (g~cm$^{-2}$)   \\
\hline
$p$ & W98          & $3.4\times 10^{-18}$ & 44              \\
$p$ & M02          & $5.4\times 10^{-18}$ & 38              \\
\hline
$e$   & C00          & $3.3\times 10^{-20}$ & 71              \\
$e$   & E00          & $4.9\times 10^{-18}$ & 35              \\
\hline
\end{tabular}
\label{tabchi}
\end{center}
\end{table}

\begin{figure}[t]
\begin{center}
\resizebox{\hsize}{!}{\includegraphics{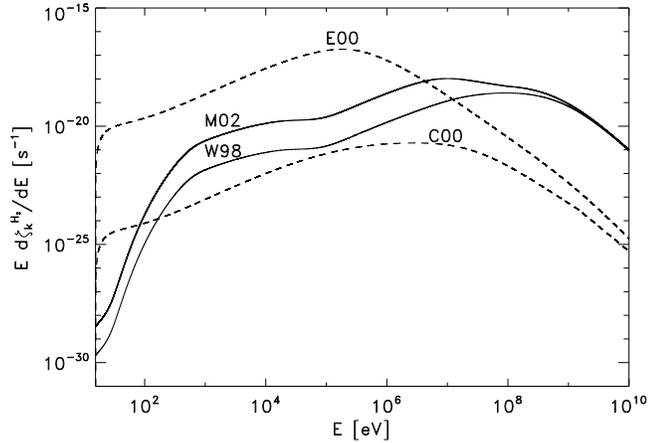}}
\caption[]{Differential contribution to the ionization rate $E{\rm d}\zeta_k^{{\rm H}_2}/{\rm d}E$ 
per logarithmic interval of kinetic energy, for the four spectra considered
in this paper at a depth $N({\rm H}_2)=10^{22}$~cm$^{-2}$ ({\em solid curves}, protons; 
{\em dashed curves}, electrons).}
\label{Figure14}
\end{center}
\end{figure}

It is important to stress that a large contribution to the
ionization of H$_2$ comes from low-energy protons and electrons
constantly produced (in our steady-state model) by the slowing-down of
more energetic particles loosing energy by interaction with the ambient
H$_2$. In Fig.~\ref{Figure14} we show the differential contribution of CRs
protons and electrons to the ionization rate at a depth of $N({\rm
H}_2)=10^{22}$~cm$^{-2}$, corresponding to the typical column density
of a dense cloud. For protons and heavy nuclei, the bulk of the
ionization is provided by CR in the range 1~MeV--1~GeV and  by a
``shoulder'' in the range 1--100~keV produced by slowed-down protons.
This low-energy tail is produced during the propagation of CR protons
in the cloud even when the incident spectrum is devoid of low-energy
particles (as shown in Fig.~\ref{Figure9} for the W98 spectrum).  The
largest contribution of CR-electrons to the ionization is distributed
over energies in the range 10~keV--10~MeV, again reflecting the
distribution of electrons in the propagated spectra (see
Fig.~\ref{Figure11} and \ref{Figure12}).  Thus, the ionization rate at any
depth in a cloud cannot be calculated by simply removing from the
incident spectrum particles with energies corresponding to ranges below
the assumed depth.

\section{Comparison with observations}
\label{comparison}

\begin{figure}[t]
\begin{center}
\resizebox{\hsize}{!}{\includegraphics{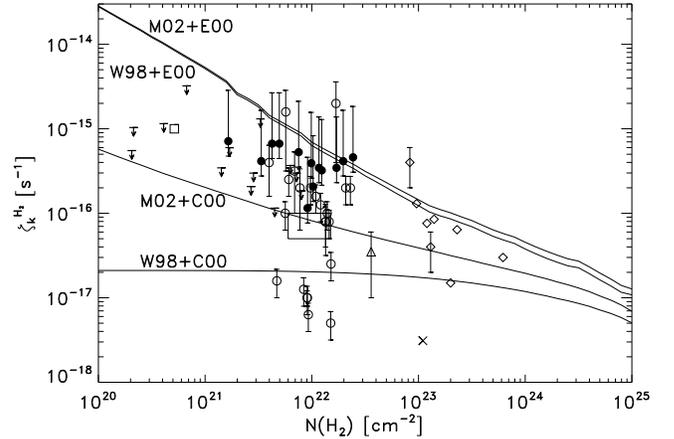}}
\caption[]{Total CR ionization rate $\zeta^{{\rm H}_2}$ a function of
$N({\rm H}_2)$ according to our models ({\em solid} curves). Observational
data: 
{\em filled circles} and {\em upper limits}, diffuse clouds (Indriolo et al.~2007); 
{\empty square}, diffuse cloud towards $\zeta$~Per (Shaw et al.~2008);
{\em empty circles}, dense cores (Caselli et al.~1998); {\empty triangle}, prestellar
core B68 (Maret \& Bergin~2007);
{\em filled squares}, protostellar envelopes (see Table~\ref{mass_env} 
for references); ({\em cross} massive star-forming region DR21(OH) (Hezareh et al.2008). 
The {\em box} indicates the range of column densities and CR ionization rates compatible 
with the data analyzed by Williams et al.~(1998).} 
\label{Figure15}
\end{center}
\end{figure}

To obtain the total CR ionization rate in molecular clouds, we sum the
ionization rates of protons (corrected for heavy nuclei as in
Appendix~\ref{approx_heavy}) and electrons. With two possible spectra
for each component, we obtain four possible profiles of $\zeta^{{\rm
H}_2}$. These are shown in Fig.~\ref{Figure15} as function of $N({\rm
H}_2)$, compared with a compilation of empirical determinations of
$\zeta^{{\rm H}_2}$ in diffuse and dense clouds. Our data sample
includes:  ({\em i}\/) diffuse clouds with $N({\rm H}_2)$ from
$10^{20}$~cm$^{-2}$ to $10^{22}$~cm$^{-2}$ (14 detections and 15 upper
limits, from Indriolo et al.~2007, including previous data of McCall et
al.~2002) and for the $\zeta$~Per line-of-sight (Shaw et al.~2008);
({\em ii}\/) molecular cloud cores with $N({\rm H}_2)$ from
$10^{21}$~cm$^{-2}$ to $10^{22}$~cm$^{-2}$ (data for low-mass cores
from Caselli et al.~1998, Williams et al.~1998, and for the prestellar
core B68 from Maret \& Bergin~2007); ({\em iii}\/) massive protostellar
envelopes with $N({\rm H}_2)$ from $10^{22}$~cm$^{-2}$ to
$10^{23}$~cm$^{-2}$ (see Table~\ref{mass_env} and references therein).

\begin{table}[t]
\begin{center}
\caption{CR ionization rate toward massive protostellar envelopes}
\begin{tabular}{llll}
\hline
    & $N({\rm H}_2)$ & $\zeta^{{\rm H}_2}$ \\
    & (cm$^{-2}$)    &  (s$^{-1}$)         \\
\hline
NGC~2264~IRS    & $8.3\times 10^{22}(^{\rm a})$  & $(4\pm 2)\times 10^{-16}(^{\rm a})$  \\  
GL~2136         & $1.2\times 10^{23}(^{\rm b})$  & $7.6\times 10^{-17}(^{\rm c})$       \\
W3~IRS5         & $1.3\times 10^{23}(^{\rm a})$  & $(4\pm 2)\times 10^{-17}(^{\rm a})$  \\  
GL~2591         & $9.6\times 10^{22}(^{\rm d})$  & $1.3\times 10^{-16}(^{\rm c})$       \\
GL~490          & $2.0\times 10^{23}(^{\rm b})$  & $1.5\times 10^{-17}(^{\rm c})$       \\
W~33A           & $6.2\times 10^{23}(^{\rm b})$  & $3.0\times 10^{-17}(^{\rm c})$       \\
W3~IRS5         & $2.3\times 10^{23}(^{\rm b})$  & $6.4\times 10^{-17}(^{\rm c})$       \\
S~140           & $1.4\times 10^{23}(^{\rm b})$  & $8.5\times 10^{-17}(^{\rm c})$       \\
DR21(OH)        & $1.1\times 10^{23}(^{\rm e})$  & $3.1\times 10^{-18}(^{\rm e})$       \\
\hline
\end{tabular}
\label{mass_env}
\end{center}
\tiny{$(^{\rm a})$ from de Boisanger, Helmich, \& van Dishoeck~(1996)} \\
\tiny{$(^{\rm b})$ CO column density and CO/H$_2$ ratio from van der Tak et al.~(2000)} \\
\tiny{$(^{\rm c})$ CR rate from van der Tak \& van Dishoeck~(2000)} \\
\tiny{$(^{\rm d})$ from Doty et al.~(2002)} \\
\tiny{$(^{\rm e})$ from Hezareh et al.~(2008)} \\
\end{table}

The observational value of $\zeta^{{\rm H}_2}$ in diffuse clouds is
obtained from the steady-state abundance of H$_3^+$, produced by the CR
ionization of H$_2$ followed by a fast charge exchange reaction with
H$_2$ and destroyed mainly by electron recombination.  The situation
for dense molecular clouds and protostellar envelopes is more
complicated. In the dense molecular gas, H$_3^+$ is removed by
reactions with other molecules and atoms of the gas, e.g., by reaction
with CO to form HCO$^+$ and with O to form OH$^+$.  Thus $\zeta^{{\rm
H}_2}$ can be determined from the measured abundance of a variety of
molecular ions such as HCO$^+$, DCO$^+$ and N$_2$H$^+$ (see e.g Caselli
et al.~1998; van der Tak \& van Dishoeck~2000; Doty et al.~2002).  The
resulting rates are nonetheless very uncertain, as they depend on the
depletion of elemental C and O from their cosmic abundances, especially
for clouds with a low degree of ionization, and are generally sensitive
to the adopted chemical model.  Here we adopt the values of
$\zeta^{{\rm H}_2}$ derived by Caselli et al.~(1998) with the data of
Butner, Lada, \& Loren~(1995) and the chemical model of Leung, Herbst,
\& Huebner~(1984). We view the range of values of $\zeta^{{\rm H}_2}$
obtained for different depletion factors as an indication of the
associated uncertainties in the model determinations.  In contrast with
the study of Caselli et al.~(1998), Williams et al.~(1998) analyze
molecular line data for a sample of low-mass cores using the chemical
models of Bergin, Langer \& Goldsmith~(1995) and Bergin \&
Langer~(1997). They conclude that a single value (or a narrow range of
values) of $\zeta^{{\rm H}_2}$ can reproduce reasonably well the
observations for the majority of cores in their sample.

The comparison between model results and observational data shown
in Fig.~\ref{Figure15} should be taken as indicative and interpreted in a
statistical sense, as also suggested by the large spread of values of
$\zeta^{{\rm H}_2}$ at each value of $N({\rm H}_2)$. First, the {\em
observational} $N({\rm H}_2)$ is the entire column density through the
cloud, whereas the {\em model} $N({\rm H}_2)$ is the column traversed
by CRs incident over the cloud's surface. The exact relation between
the quantities depend on factors like the cloud geometry and
orientation with respect to the line-of-sight, and the variation of CR
ionization rate with depth within the cloud. In addition, for the
cloud cores of Caselli et al.~(1998) we adopted the H$_2$ column
density estimated by Butner et al.~(1995) from measurements of
C$^{18}$O multiplied by a factor of 2, to account for depletion of CO
onto grains (Caselli et al.~1998). In fact, at the time of the study by
Caselli et al.~1998, the almost complete disappearance of CO from the
gas phase {\it in cloud cores} was still unknown. Second, many of
the sight-lines where $\zeta^{{\rm H}_2}$ has been determined in
diffuse clouds may have multiple cloud components, which would reduce
the column density of a single cloud. It is probably safe to conclude that
the {\em observational} column density is an {\em upper limit} to the
column density traversed by CRs incident on each cloud, and therefore
the data shown in Fig.~\ref{Figure15} should probably be shifted to the
left by a factor or 2 or so.  We will address the problems relative to
cloud geometry and the effects of magnetic fields in a
subsequent work.  At any rate, from the comparison with observational
data, shown in Fig.~\ref{Figure15}, we can draw the following
conclusions:

({\em i}\/) Although the gas column density of the object is by no
means the only parameter controlling the CR ionization rate, the data
suggest a decreasing trend of $\zeta^{{\rm H}_2}$ with increasing
$N({\rm H}_2)$, compatible with our models M02+C00, W98+E00, W98+C00.
However, the measured values of $\zeta^{{\rm H}_2}$ are very uncertain,
especially in dense environments. Part of the large spread in the
sample of cloud cores may be due to a poor understanding of the
chemistry.

({\em ii}\/) The highest values of $\zeta^{{\rm H}_2}$, measured in
diffuse clouds sight lines, could be explained if CR electrons are
characterized by a rising spectrum with decreasing energy. The E00
spectrum represents an extreme example of this kind, and it results in
values of $\zeta^{{\rm H}_2}$ somewhat in excess of the diffuse clouds
observations. The same spectrum accounts simultaneously for the CR
ionization rates measured in most protostellar envelopes of much higher
column density.  Conversely, a spectrum of protons and heavy nuclei
rising with decreasing energy, like the M02 spectrum, can provide alone
a reasonable lower limit for the CR ionization rate measured in diffuse
clouds.

({\em iii}\/) Without a significant low-energy (below $\sim 100$~MeV)
component of electrons and/or protons and heavy nuclei, it is
impossible to reproduce the large majority of observations. The
combination of the C00 spectrum for electrons with the W98 spectrum for
protons and heavy nuclei clearly fails over the entire range of column
densities.  Finally, a few molecular cloud cores and one dense envelope
characterized by $\zeta^{{\rm H}_2} \le 10^{-17}$~s$^{-1}$ can only be
explained by invoking the CR suppression mechanisms mentioned in
Sect.~\ref{introduction} not considered in this work.

In a recent paper, published after our work was completed,
Indriolo, Fields \& McCall~(2009) analyze the implications of the CR
ionization rate measured with H$_3^+$ in diffuse and dense clouds with
an approach similar to that adopted in this paper. As in the present
work, Indriolo et al.~(2009) reach the conclusion that a low-energy CR
component, likely produced by weak, localized shocks, can account for
the relatively high CR ionization rate measured in diffuse clouds.
However, their ``best-fitting'' CR-proton spectrum increases below
$E_p\approx 100$~MeV as $E_p^{-2.15}$, much more steeply than our
steepest spectrum (M02, increasing as $E_p^{-1}$). Indriolo et al.
avoid a too large CR ionization (and heating) rate by cutting off the
proton spectrum below $E_p=2$~MeV and $E_p=10$~MeV for diffuse and
dense clouds, respectively. These energies correspond roughly to proton
ranges in H$_2$ of $10^{21}$~cm$^{-2}$ and $10^{22}$~cm$^{-2}$,
respectively. However, our work shows instead that when CR propagation
is properly taken into account, low-energy particles are continuously
produced by more energetic particles when they slow-down by interacting
with the ambient medium (see Sect.~\ref{ionization}), making
significantly to the ionization rate. In addition, we found that the
contribution of CR-electrons to the ionization of H$_2$, neglected by
Indriolo et al.~(2009), can be significant or even dominant over the
contribution of CR-protons and heavy nuclei, without violating the
available observational constraints.

\section{Conclusions}
\label{conclusions}

The comparison between our models and the observational data available
for diffuse clouds, dense cores and massive protostellar envelopes
indicates that good agreement between theory and observations can be
obtained for the CR ionization rate of the ISM by including CR
electrons with an energy spectrum increasing towards low energies, as
also suggested by Webber~(1998).  In order to draw more stringent
conclusions, it is necessary to use observational constraints derived
from the ionization rates of diffuse clouds, as seen in
Fig.~\ref{Figure15} where the models differ mainly at low column
densities. Our study points out the current limits towards a more
accurate understanding of the ionization due to cosmic rays. There are
essentially two limits: (1) the uncertainty in the CR spectrum at
energies below $\sim 1$~GeV, and (2) the uncertainties in the
empirically determined values of $\zeta^{{\rm H}_2}$ in diffuse and
dense molecular clouds.  The estimates of the ionization rate depend
sensitively on the complex set of chemical reactions governing the
chemistry of the ISM, particularly on the so-called depletion processes
that transfer molecules and ions from the gas phase to the solid phase.
Despite these observational uncertainties, several important
conclusions clearly emerge from our study:

({\em i}\/) values of $\zeta^{{\rm H}_2}$ measured in diffuse clouds
are greater on average by an order of magnitude than those ones
measured in dense molecular clouds. If confirmed, these data imply the
presence of a CR proton and/or CR electron spectrum which increases at
low energies. Thus, a combination of the spectra W98 and C00 for
protons and electrons, respectively, is excluded by this set of
observations;

({\em ii}\/) values of $\zeta^{{\rm H}_2}$ measured in dense molecular
clouds span a range of about two orders of magnitude and are subject to
considerable uncertainty. It is difficult to establish how much of the
observed spread is due to variations in the CR ionization rate. It is
likely that in dense clouds the effects of magnetic fields on the
propagation of CR particles cannot be neglected. In addition, it might
be necessary to take into account the density distribution inside each
cloud;

({\em iii}\/) the values of $\zeta^{{\rm H}_2}$ measured in massive
protostellar envelopes are somewhat higher than the predictions of our
models at the corresponding column densities. This seems to suggest the
presence of further ionization sources in these objects, as, for
example, X-ray emission from the young stellar objects;

({\em iv}\/) The exponential attenuation of the CR ionization rate
assumed in many studies is only established for column densities larger
than $\sim 10^{25}$~cm$^{-2}$. For the lower column densities
considered in this work, the ionization rate decreases as $\zeta^{{\rm
H}_2}\propto N({\rm H}_2)^{-a}$ with $a\approx 0.4$--0.8 for the
spectra that best reproduce the observational data.

It is reasonable to suppose that some of the observational
uncertainties discussed here will be removed or reduced in the near
future. With respect to the calculation of the ionization rate in dense
clouds, the understanding of the complex chemical reactions and of the
depletion processes appear to be improving rapidly. Regarding the
extrapolation of the measured CR spectra to low energies, the particle
fluxes measured by Voyager and Pioneer spacecrafts outside of the solar
magnetopause should provide important constraints on the energy
distribution of CR protons and CR electrons, and, in any case, improve
our understanding of solar modulation.

\acknowledgements
MP and DG acknowledge support from the Marie-Curie Research Training
Network ``Constellation'' (MRTN-CT-2006-035890). AEG acknowledges
support from NSF NSF grant AST-0507423 and NASA grant NNG06GF88G. We thank an anonymous referee for insightful comments that helped to
improve our paper.

\appendix
\section{Approximated corrections for heavy-nuclei}
\label{approx_heavy}

In the Bethe-Born approximation, the cross section for the collisional
ionization of an atom or molecule depends only on the charge $Z_k$ and
the velocity $v_k$ of the incident particle. If $A_k$ is the number of
nucleons in the incident particle, the ionization cross section is
$\sigma_k(E_k)\approx Z_k^2\sigma(\epsilon)$, where $\epsilon=E_k/A_k$
is the kinetic energy per nucleon and $\sigma(\epsilon)$ is the same
for all particle's species. If, in addition, the spectra of CR protons
and heavy nuclei can be approximately described by a single function
$j(\epsilon)$ such that $j_k(E_k)\,\ud E_k \approx f_k j(\epsilon)\,\ud
\epsilon$, with $f_k$ representing the fractional abundance by number
of species $k$, it is possible to reduce the calculation of the
ionization rate by heavy-nuclei impact to that of protons, as
\begin{eqnarray}
\lefteqn{
\sum_{k\ge 1}\int_{I({\rm H}_2)}^{E_{\rm max}} 
j_k(E_k)[1+\phi_k(E_k)]\sigma_k^{\rm ion.}(E_k)\,\ud E_k\approx} \nonumber \\
& & 
(1+\eta)\int_{I({\rm H}_2)}^{E_{\rm max}} j_p(E_p)[1+\phi_p(E_p)]
\sigma_p^{\rm ion.}(E_p)\,\ud E_p,
\end{eqnarray}
where $\eta$ is the correction factor for heavy nuclei ionization,
\be
\eta\equiv\sum_{k\ge 2}\frac{f_k}{f_p}Z_k^2.
\ee
Similarly, the correction factor to account for the energy density
of heavy nuclei (eq.~\ref{en_den}) is given by 
\be
\xi\equiv\sum_{k\ge 2} \frac{f_k}{f_p}A_k.
\ee
Assuming for the CR abundance of heavy nuclei the standard solar
abundance (Anders \& Grevesse~1989), we obtain $\eta=0.51$, and
$\xi=0.41$, in agreement with the values $\eta=0.50$ and
$\xi=0.42$ of Indriolo et al.~(2009).

\clearpage

\end{document}